\documentstyle[aaspp4,emulateapj5,apjfonts]{article}

\gdef\h50min{$h_{50}^{-1}$}
\gdef\h70min{$h_{70}^{-1}$}
\gdef\kms{km\,s$^{-1}$}
\gdef\sbunit{mag\,arcsec$^{-2}$}

\lefthead{}
\righthead{}
\slugcomment{Astronomical Journal, in press (December 2005 issue)}
\begin{document}

\title{The Recent and Continuing Assembly of Field Ellipticals by
Red Mergers}
\author{
Pieter G.\ van Dokkum
\affil{Department of Astronomy, Yale
University, New Haven, CT 06520-8101; dokkum@astro.yale.edu}
}

\begin{abstract}

We present a study of tidal debris associated with 126 nearby red
galaxies, selected from the 1.2
degree$^2$ Multiwavelength Survey by Yale-Chile (MUSYC) and the
9.3 degree$^2$ NOAO Deep Wide-Field Survey.  In the full sample
67 galaxies
(53\,\%) show morphological signatures of
tidal interactions, consisting of
broad fans of stars, tails, and other  asymmetries at very faint
surface brightness levels.
When restricting the sample to the
86 bulge-dominated early-type
galaxies  the fraction
of tidally disturbed galaxies rises to 71\,\%, which implies that for
every ``normal'' undisturbed elliptical there are two which show
clear signs of interactions.
The tidal features are red and smooth, and
often extend over $>50$\,kpc. Of the tidally distorted
galaxies about $2/3$ are remnants and $1/3$ are interacting
with a companion galaxy.
The companions are usually bright red
galaxies as well: the median $R$-band luminosity
ratio of the tidal pairs is 0.31, and the median color difference
after correcting for the slope of the color-magnitude relation is
$-0.02$ in $B-R$. If the ongoing mergers are representative for the
progenitors of the remnants $\sim 35$\,\% of bulge-dominated
galaxies experienced a merger with mass ratio $>1:4$ in
the recent past. With further assumptions it is estimated
that the present-day mass
accretion rate of galaxies on the red sequence $\Delta M/M = 0.09 \pm
0.04$\,Gyr$^{-1}$. For a constant or
increasing mass accretion rate with redshift, we find that
red mergers may lead to an
evolution of a factor of $\gtrsim 2$ in
the stellar mass density in luminous red galaxies over the redshift range
$0<z<1$, consistent with recent studies of the evolution of the
luminosity density.
We conclude that most of today's field elliptical galaxies
were assembled at low redshift
through mergers of gas-poor,
bulge-dominated systems. These ``dry'' mergers are
consistent with the high central
densities of ellipticals, their
old stellar populations, and the strong correlations
of their properties. It will be interesting to determine whether
this mode of merging only plays an important role at low redshift
or is relevant
for galaxies at any redshift if they exceed a critical mass scale.

\end{abstract}


\section{Introduction}

\begin{figure*}[b]
\epsfxsize=17.5cm
\epsffile[54 314 533 513]{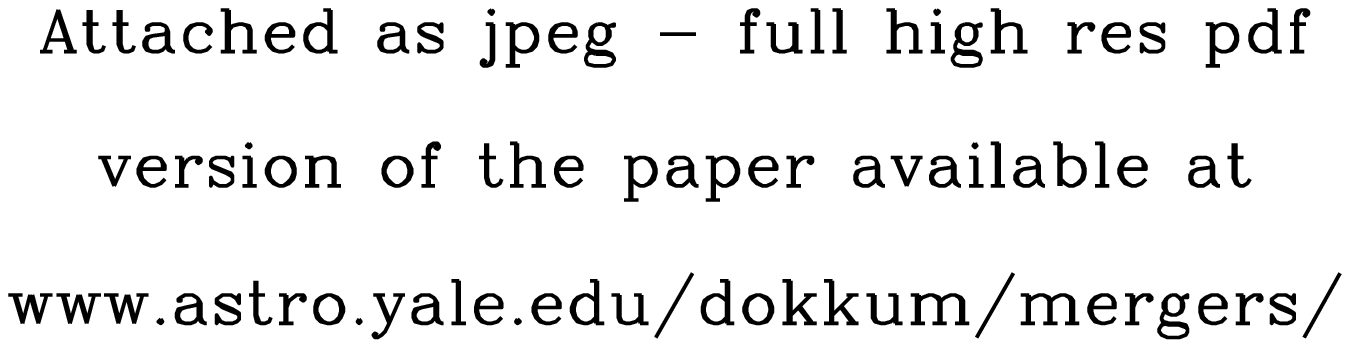}
\caption{\small
Illustration of an off-axis collision
between hot stellar systems, with
mass ratio 3:1. The simulation was performed with an implementation
of the Barnes \& Hut (1986) tree code, using 65,536 particles.
The galaxies merge quickly; after the merger
the only morphological signature is the presence of a broad
asymmetric fan of material, which rapidly becomes more diffuse.
Note the lack of tidal tails, which only form if there is
a significant cold component.
\label{sim.plot}}
\end{figure*}

Although greatly outnumbered by spiral galaxies, elliptical galaxies
contain $\sim 20$\,\% of the stellar mass in the present-day Universe
({Fukugita}, {Hogan}, \&  {Peebles} 1998)
and dominate the high mass end of the galaxy mass function
(e.g., {Nakamura} {et~al.} 2003). Their
status at the top of the food chain is a strong motivation for
their study as it provides insight in the processes governing
the build-up of galaxies over cosmic time, the nature of the most
luminous galaxies at $z\sim 3$ and beyond, the formation of the
first stars, and the origins of supermassive black holes.

Despite vigorous research efforts it is still not known how and when
elliptical galaxies were formed.  Separating the time of the formation
of their {\em stars} from the time of their {\em assembly}, a partial answer is
that most of the stars in the most massive ellipticals were formed at
high redshift. This result is supported by a large number of studies
of nearby and distant galaxies, and applies to ellipticals in the
general field as well as those in rich clusters
(e.g., {Bower}, {Lucey}, \& {Ellis} 1992; {Ellis} {et~al.} 1997; {Bernardi} {et~al.} 1998, 2003; {van Dokkum} {et~al.} 1998a, 1998b; {Treu} {et~al.} 1999, 2002; {van Dokkum} {et~al.} 2001a; {Holden} {et~al.} 2005).
Even when accounting for possible selection effects due to
morphological evolution (``progenitor bias''; {van Dokkum} \& {Franx} 2001),
it is very difficult
to fit the observed evolution of massive ellipticals with
formation redshifts substantially lower than $z\sim 2$.
Although there is general agreement that the mean age is high, there are
strong indications that not all stars in all elliptical
galaxies formed at high redshift: Trager et al.\ (2000) and others
(e.g., {Yi} {et~al.} 2005) find evidence for a ``sprinkling'' of recent star
formation in otherwise old ellipticals, and recent studies of the
evolution of the Fundamental Plane relation ({Djorgovski} \& {Davis} 1987) find
evidence that low mass early-type galaxies have lower
luminosity-weighted ages than high mass ones (Treu et al.\ 2005; van
der Wel et al.\ 2005). 

Much less is known about the assembly of elliptical galaxies,
that is, when and how the galaxies were ``put together''.  In the
literature, a distinction is often made between models in which
elliptical galaxies form from the collapse of primordial gas clouds
(``monolithic collapse''; e.g.,
{Eggen}, {Lynden-Bell}, \&  {Sandage} 1962; {Jimenez} {et~al.} 1999), or in
mergers of smaller galaxies (``hierarchically'';
e.g. {Toomre} \& {Toomre} 1972; {White} \& {Frenk} 1991).  Within the framework of a merger
model the question of assembly can be specified as, first, the typical
age of the last major merger; and second, the nature of the progenitor
galaxies.
In many individual cases we can readily answer both
questions: abundant evidence leaves little doubt that some ellipticals
formed recently through the merger of spiral galaxies, or will do so
in the near future (e.g., {Toomre} \& {Toomre} 1972; {Schweizer} 1982; {Hibbard} \& {van Gorkom} 1996).
However, it is not clear
whether such mergers are
exceptional or responsible for the formation of the
majority of ellipticals.

Among the observational
arguments {\em for} recent mergers are the detection
of ``fine-structure'' (e.g., shells and ripples) in a large number
of nearby ellipticals ({Malin} \& {Carter} 1983;
Schweizer et al.\ 1990); the presence of
kinematically-decoupled cores ({Franx} \& {Illingworth} 1988; {Bender} 1988);
small-scale dust seen in the majority of ellipticals
(e.g., {van Dokkum} \& {Franx} 1995); (some) studies of the redshift evolution
of galaxy morphologies, and of close pairs (e.g., {Le F{\` e}vre} {et~al.} 2000; {Patton} {et~al.} 2002; {Conselice} {et~al.} 2003); and strong evolution of the mass density of
red galaxies to $z\sim 1$, inferred from the COMBO-17 survey
({Bell} {et~al.} 2004). Furthermore, and perhaps most importantly,
the hierarchical assembly of galaxies
is a central aspect of current models for galaxy formation
in $\Lambda$CDM cosmologies (e.g., {Kauffmann}, {White}, \&  {Guiderdoni} 1993; {Cole} {et~al.} 2000; {Meza} {et~al.} 2003).

Among arguments {\em against} the merger hypothesis are the
low probability of mergers in
virialized clusters, which are dominated
by ellipticals ({Ostriker} 1980; {Makino} \& {Hut} 1997); the fact that
the central densities
of ellipticals are too high to be the result of dissipationless
mergers of pure disks ({Carlberg} 1986; {Hernquist} 1992);
the low scatter
in the color-magnitude relation and Fundamental Plane,
which is inconsistent with
the diversity in stellar populations expected from spiral
galaxy mergers
(e.g., {Bower}, {Kodama}, \& {Terlevich} 1998, Peebles 2002);
the much higher specific frequency of globular clusters
for ellipticals than for
spirals (although some globulars may be formed during mergers;
e.g., {Schweizer} {et~al.} 1996); the observation that known remnants of
spiral galaxy mergers have underluminous X-ray halos
({Sansom}, {Hibbard}, \&  {Schweizer} 2000);
the existence of the $M_{\bullet} - \sigma$
relation ({Gebhardt} {et~al.} 2000; {Ferrarese} \& {Merritt} 2000); and the presence
of massive galaxies at high redshift (e.g., {Franx} {et~al.} 2003; {Daddi} {et~al.} 2004; {Glazebrook} {et~al.} 2004)

These apparently contradictory lines of evidence may be largely reconciled
by postulating that most elliptical galaxies formed through (nearly)
dissipationless (or ``dry'')
mergers of red, bulge-dominated galaxies rather than mergers of spiral disks.
Motivated by the
comparatively red colors of ellipticals exhibiting fine-structure
{Schweizer} \& {Seitzer} (1992) discuss this possibility,
but argue that mergers between early-type galaxies are
statistically unlikely as the ``median'' field galaxy
is an Sb spiral (see also Silva \& Bothun 1998).
However, most mergers likely occur in
groups, where the early-type galaxy
fraction is much
higher than in the general field (e.g., {Zabludoff} \& {Mulchaey} 1998).
Furthermore, semianalyical models of galaxy formation have predicted that
the most recent mergers of bright ellipticals were
between gas-poor,
bulge-dominated galaxies ({Kauffmann} \& {Haehnelt} 2000; {Khochfar} \& {Burkert} 2003), and several
studies have shown that such merging is consistent with the
observed Fundamental Plane relation (e.g., {Gonz{\' a}lez-Garc{\'{\i}}a} \& {van Albada} 2003; {Boylan-Kolchin}, {Ma}, \&  {Quataert} 2005). Finally, mergers between red galaxies may be common
in young, unvirialized galaxy clusters at $z\sim 1$
({van Dokkum} {et~al.} 1999, 2001b, Tran et al.\ 2005).

Although some early-type/early-type mergers are known to
exist in the local universe
(e.g., {Davoust} \& {Prugniel} 1988; {Combes} {et~al.} 1995), little is known about their frequency.
Here we investigate the relevance of dry merging by
analyzing the frequency and nature of tidal distortions associated
with a well-defined sample of bright red galaxies.
This study was motivated by the results of {Kauffmann} \& {Haehnelt} (2000),
{Khochfar} \& {Burkert} (2003),
and {Bell} {et~al.} (2004),
the observations of red mergers in
$z \sim 1$ clusters, and the advent of very deep photometric
surveys over wide areas. We assume $\Omega_m=0.3$,
$\Omega_{\Lambda}=0.7$, and $H_0=70$\,\kms\,Mpc$^{-1}$ (Spergel
et al.\ 2003).

\section{Morphological signatures}
\label{signatures.sec}

Mergers between gas-poor, bulge-dominated galaxies are generally
more difficult to recognize than mergers between gas-rich disks.
Elliptical-elliptical mergers do not develop prominent tidal
tails dotted with star forming regions, but are instead characterized by
the ejection
of broad ``fans'' of stars, and in certain cases an asymmetric deformation
of the inner isophotes  (e.g., {Rix} \& {White} 1989; {Balcells} \& {Quinn} 1990; {Combes} {et~al.} 1995).
Tails may develop if one of the progenitors rotates or has a disk
component ({Combes} {et~al.} 1995), but as there is no cold, young component
these are expected to be more diffuse than those seen in
encounters between late-type galaxies.

These effects are illustrated in
Fig.\ \ref{sim.plot}, which shows the evolution of an off-axis collision
between two hot stellar systems. The simulation was performed with
an implementation of
the {Barnes} \& {Hut} (1986) hierarchical tree code, using 65,536 particles.
The galaxies have a 3:1 mass ratio
and are represented by Plummer models. The
simulation illustrates the rapid merging of the two bodies, the lack
of tidal tails, and the development of a large fan of material. The
central parts of the merger remnant quickly settle in a new
equilibrium configuration, whereas the material in the outer parts --
where dynamical timescales are long --
gradually becomes more diffuse.  For a more
general analysis of the development of elliptical-elliptical mergers,
including the effects of rotation and different mass ratios, we
refer to {Combes} {et~al.} (1995) and subsequent papers.

The surface brightness of tidal features associated with
this type of mergers is
comparatively low: the features consist
of old, red stellar populations, and have higher $M/L$ ratios
than the blue tails associated with
mergers of late-type galaxies. Furthermore, they are
typically more diffuse, due to the lack of a cold
disk in the progenitors. After the merger,
the evidence of the past encounter
becomes increasingly difficult to detect as dynamical evolution
reduces the surface brightness further (e.g., {Mihos} 1995).
No detailed simulations have been done of the expected surface
brightness evolution of tidal features resulting from
early-type mergers, but a very rough
estimate can be obtained from the simulation shown in Fig.\
\ref{sim.plot}. Calibrating the output such that
the average surface brightness within the effective
radius of the remnant
$\langle \mu_R \rangle = 20$\,mag\,arcsec$^{-2}$
({J\o{}rgensen}, {Franx}, \&  {Kj\ae{}rgaard} 1995), we find that
the ``fan'' of material to the upper left has $\mu_R \sim
25$ at $t=10$, i.e., immediately after
the merger.  The following timesteps show
a continued rapid decrease of the brightness, reaching
$\mu_R \sim 27$ at $t=15$,
the final time step of the simulation.

It is difficult to reach such low surface brightness levels
observationally:
extreme enhancements of
photographic plates enabled {Malin} \& {Carter} (1983) to detect features to
$\mu \sim 26.5$,
similar to the limits reached by {Schweizer} \& {Seitzer} (1992) with their
deep CCD exposures on the Kitt Peak 0.9\,m telescope.
Detecting features with surface brightness $27-28$ requires
several hours of integration time on 4\,m class telescopes, and
to image a sizeable
sample of nearby galaxies would require hundreds of hours. Furthermore, flat
fielding would have to be accurate to $\lesssim 0.1$\,\%
over the entire $10'-30'$ field.

These practical difficulties can be solved by exploiting the
fact that the $(1+z)^4$ cosmological
surface brightness dimming is a very slow function of redshift
for $z\ll 1$. Whereas moving a galaxy from $z=0.01$ to
$z=0.1$ reduces its total brightness by
5 magnitudes, its surface brightness changes by only 0.35 mag.
In practice it is easier to detect very faint surface
brightness features around a galaxy at $z=0.1$ than at
$z=0.01$
as it occupies a $70 \times$ smaller area of
the detector. Furthermore, and most importantly,
a study of tidal features
at $z\sim 0.1$ does not require a dedicated survey but
can be done with existing imaging data.

\section{Data}

We use data from two deep extra-galactic surveys: the Multiwavelength
Survey by Yale-Chile (MUSYC; Gawiser et al.\ 2005), and the NOAO
Deep Wide-Field Survey (NDWFS; Januzzi \& Dey 1999,
Januzzi et al.\ 2005, Dey et al.\ 2005).
The data products from the optical imaging components of the two
survey are quite similar, as they both use the $8192 \times 8192$ pixel
MOSAIC cameras on the NOAO 4m telescopes. The full dataset
covers 10.5 square degrees of sky, reaching $1\sigma$
surface brightness levels of $\mu \approx 29$\,\sbunit.

\subsection{MUSYC}

MUSYC is a deep optical/near-infrared imaging and
spectroscopic survey of two Southern and two equatorial $30' \times 30'$
fields, chosen for their low background and the availability
of data at other wavelengths. The fields are the
Extended Chandra Deep Field South (E-CDFS), the
Extended Hubble Deep Field South (E-HDFS), the field centered
on the $z=6.3$ Sloan Digital Sky Survey quasar SDSS\,1030+05
({Becker} {et~al.} 2001), and the blank low background field
CW\,1255+01. The primary goal of the survey is the study of
normal and active galaxies at $z>2$.

The survey design and data analysis procedures
are described in Gawiser et al.\ (2005). Briefly,
the optical imaging data consist of deep $UBVRIz$ exposures
obtained with MOSAIC-I on the Mayall 4m telescope
on Kitt Peak and MOSAIC-II on the Blanco 4m telescope on
Cerro Tololo. The images have a scale of $0\farcs 267$\,pix$^{-1}$,
and typical seeing of $1\farcs 1$.
The data reach depths of $\approx 26$ in $U$,
$\approx 26.5$ in $B$, $V$, and $R$, $\approx 25$ in $I$, and $\approx 24$
in $z$ (AB magnitudes, $5\sigma$
point source detections), with some variation between the
fields. The full exposure time is typically realized
over an area of $33' \times 33'$, and the total optical survey
area is $1.2$\,deg$^2$.

For detection of low surface brightness features we used the
co-added ``BVR'' images (see Gawiser et al.\ 2005), which
are combinations of the $B$, $V$, and $R$ frames. The total
effective integration time of
these frames is typically $\sim 8$ hours. The final depth
reaches magnitude $\approx 27$ (AB; $5\sigma$ point source
detection), with slight variations from field to field.

\subsection{NOAO Deep Wide-Field Survey}

The NDWFS is a public optical and near-infrared imaging survey over
two 9.3\,deg$^2$ fields: a $3{\arcdeg} \times 3{\arcdeg}$
field in Bo\"otes, centered at 14$^h$32$^m$, $+34{\arcdeg}17\arcmin$,
and a $2.3\arcdeg{} \times 4{\arcdeg}$ field in Cetus,
centered at $2^h$10$^m$, $-4\arcdeg 30\arcmin$
(see Januzzi \& Dey 1999; Januzzi et al.\ 2005).
The primary goal of the survey is to study large scale structure
at $z>1$. We use the optical images available in
the third data release (DR3; October 22 2004). This release
comprises the entire $9.3$\,deg$^2$ Bo\"otes field.

Compared to MUSYC, the NDWFS sacrifices multi-band coverage for
area. Only three optical filters are used: $B_w$, $R$, and $I$.  The
$B_w$ filter was designed specifically for the NDWFS: it is a very
broad blue filter, effectively a combination of the standard
$U$ and $B$ filters. Exposure times vary across the field;
the median values are
8400\,s in $B_w$, 6000\,s in $R$, and $11400$\,s in
$I$. Details of DR3 can be found on the
NDWFS webpages\footnote{http://www.noao.edu/noao/noaodeep/DR3/}.

The data were not pasted together in a mosaic by NOAO
but released as 27 partially overlapping $35' \times 35'$ MOSAIC pointings.
We obtained all 27 pointings in
$B_w$, $R$, and $I$ from the NOAO archive. In order to increase
the signal-to-noise (S/N) ratio we created co-added ``BRI'' images, in the
following way. First, all images were normalized such that 1
count corresponds to AB$ = 30$\,mag. Next, the summed BRI images
were created from the $B$, $R$, and $I$ frames according to
BRI$ = B + R + 0.5 \times I$. This procedure optimizes the
S/N ratio for objects with the approximate SED of nearby elliptical
galaxies.

\subsection{Depth of Summed Images}

The detection of diffuse tidal features depends on the
limiting surface brightness level of the data, not on the point
source detection limit.
To assess the (AB) surface brightness limits
we placed random apertures with an area of
$1\arcsec$ in empty regions, and determined the rms
fluctuations. The co-added images were used, i.e., the
``BVR'' images for MUSYC and the ``BRI'' images for
NDWFS.

\begin{figure*}[t]
\epsfxsize=17cm
\epsffile[44 422 543 690]{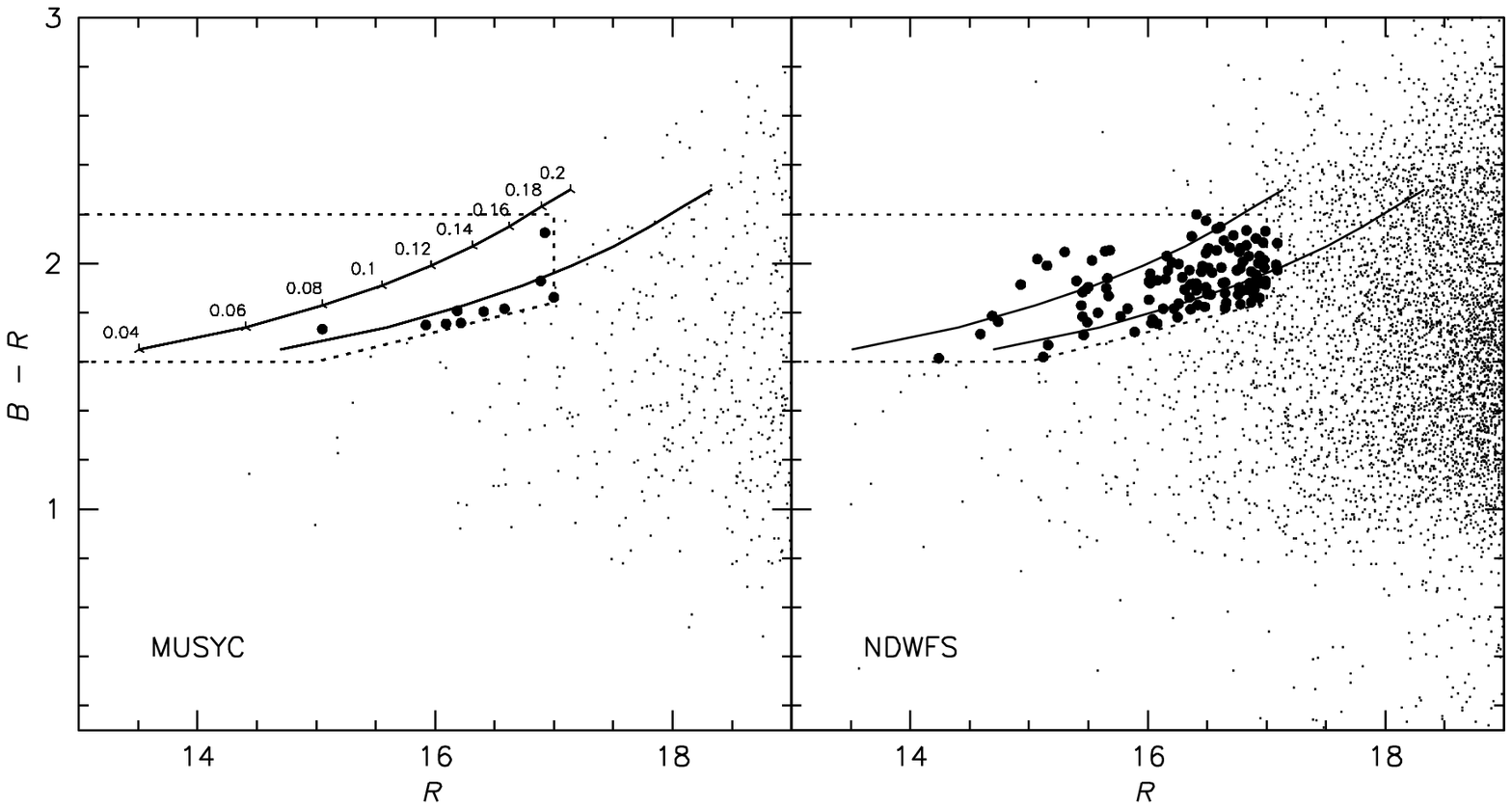}
\caption{\small
Sample selection. Solid lines show the expected $B-R$ colors and
$R$ magnitudes for $L_*$ and $3L_*$ elliptical galaxies at
redshifts $0.04\leq z\leq 0.20$. The dashed lines shows the
selection region in color and magnitude. MUSYC consists of
four 0.3\,deg$^2$ fields, including the areas
around the Chandra Deep Field
South and the Hubble Deep Field South. The NDWFS covers a
contiguous area of 9.3\,deg$^2$.
\label{sel.plot}}
\end{figure*}

The MUSYC images have a typical $1\sigma$
surface brightness limit of $\mu \approx 29.5$\,\sbunit\ whereas this
limit is $\approx 29$\,\sbunit\ for the NDWFS images.
These numbers are approximate:
the exact limit depends on the field, the local flat fielding
accuracy, and on the
assumed spectral energy distribution of the source. 
The greater depth of MUSYC was expected, as more
exposure time was devoted to the efficient
$B$, $V$ and $R$ filters.

The ability to detect low surface
brightness emission is influenced
by the extent of the features, confusion with neighboring objects,
the contrast with the smooth emission from the galaxy, and other
effects; in practice, we find we
can confidently identify features down
to $\mu \approx 28$\,\sbunit.

\section{Sample Selection}

The initial sample selection is based on color and
magnitude only. No morphological criteria are applied, for three
reasons: first, a simple color cut is
straightforward to reproduce; second,
as the galaxies are typically at $z\sim 0.1$ it can be difficult
to determine morphologies from ground-based data (see \S\,\ref{class.sec});
and third, the classification process itself might bias the sample
toward or against galaxies with tidal features.

\subsection{Catalogs}
\label{cat.sec}

The combination of the two surveys comprises 31 multi-band tiles,
each covering $\approx 0.3$\,deg$^{2}$.
Catalogs were created using the SExtractor software ({Bertin} \& {Arnouts} 1996).
SExtractor catalogs were available for both surveys but these
are optimized for faint point sources, not for large, nearby galaxies.
SExtractor was run with its default settings, with the following
changes. The detection threshold was set at $20\sigma$, with the
added requirement that 10 adjacent pixels are above the threshold,
and the mesh size for background subtraction was set at
400 pixels rather than the default 64 to avoid oversubtraction
of the background for large galaxies. Matched photometry in each
band was obtained by running SExtractor in
dual image mode, always using the $R$-band images for detection.

SExtractor's AUTO magnitudes are used as best estimates for
total magnitudes. Colors were initially
determined in fixed $5\arcsec$
diameter apertures. However, we find that
colors measured in fixed apertures are not well suited for
these large objects as they introduce distance-dependent
selection biases. Specifically,
when selecting on aperture color the sample invariably contains
large, presumably very nearby
spiral galaxies with blue disks and red bulges.
In the following, $B-R$ colors refer to
$B_{\rm AUTO} - R_{\rm AUTO}$, measured in dual image mode.

Finally,
the measured $B_w - R$ colors in the NDWFS are converted to the
standard Johnson system. The $B_w$ filter was created for the NDWFS
and has a bluer central
wavelength than the $B$ filter.
Using templates from {Coleman}, {Wu}, \& {Weedman} (1980) we
derive
\begin{equation}
(B-R) = 0.27 + 0.62 (B_w - R) + 0.07 (B_w-R)^2.
\end{equation}
This transformation holds to a few percent for $1<(B-R)<2.5$
and redshifts $z<0.2$, and was applied to the NDWFS photometry
prior to the sample selection.
Total magnitudes and colors are given on the Vega system
to facilitate comparisons to previous studies.

\subsection{Color Selection}

Galaxies are selected on the basis of their total $R$ magnitude
and $B-R$ color.
Figure \ref{sel.plot} shows color-magnitude diagrams for
the four MUSYC fields (left) and for the 27 NDWFS fields (right).
Only objects with a star/galaxy classification $<0.06$ are
shown (see {Bertin} \& {Arnouts} 1996). This classification is fairly robust
at the magnitudes of interest, and effectively removes both
non-saturated and saturated stars from the sample. Visual inspection
revealed that the software assigned a star/galaxy classification
$\geq 0.06$ to four galaxies in our sample; for these objects
the value was manually set to zero.
We note that the cores
of some bright galaxies can be saturated, which
influences the magnitudes and (particularly) the colors
in unpredictable ways. As this effect is difficult to quantify
and only influences a handful of objects we did not attempt a correction.
The distribution of points is similar in the two surveys.
The number of objects in the NDWFS is an order of magnitude
larger than in MUSYC, as expected from the difference in
area.

We selected galaxies with the colors and magnitudes
of $L>L_*$ early-type galaxies at
$0.05 < z < 0.2$. The two solid lines show the colors of the
Coleman et al.\ (1980) E/S0 template redshifted from $z=0.04$ to
$z=0.20$, normalized to $L_*$ and $3L_*$ (Blanton et al.\ 2001).
Dashed lines delineate the adopted selection region: $R<17$,
$1.6 \leq (B-R) \leq 2.2$, and $(B-R)>1.6+0.12 \times (R-15)$.
The galaxies in the dashed region are the brightest and reddest
in $10.5$ square degrees of sky.

\subsection{Additional Steps}

After visual inspection of the initial sample of 155 red objects
32 were removed for various
reasons. The discarded objects fall in a wide range of categories:
mis-classified stars (usually blends with other
stars or galaxies); objects on the edge of a field;
spurious objects (always near very bright stars); severely saturated
galaxies (although interesting in their own right, as they
are likely very bright active nuclei); and galaxies that have more
than one catalog listing. The latter occurs because the 27 NDWFS
tiles have some overlap; in these cases the data from the
overlapping pointings were added,
unless the S/N is much higher in one of the two.
We also removed 12 galaxies that are likely members of known
galaxy clusters. Eight are located within $8'$ of
NSC\,J142841+323859, a cluster at $z=0.127$,
and four are located within $4'$ of the poorer, more
compact cluster NSC\,J142701+341214 at $z\approx 0.2$
({Gal} {et~al.} 2003).

Finally, we inspected all galaxies with $17\leq R \leq 17.75$
to identify merging pairs (see \S\,\ref{tidalclass.sec}) that
are split into two objects by SExtractor, and whose
{\em combined} luminosities and colors would place them in
our selection region.
Three such pairs were identified, and the brightest galaxy
of each pair was added to the sample: 4-1190, 4-1975, and
22-2252.
The final sample thus consists of 126
red field galaxies in MUSYC and the third
data release of the NDWFS.

\subsection{Median Redshift}
\label{z.sec}

The median $R$ magnitude of galaxies in our sample is 16.4, and
the median $B-R$ color is 1.92. As can be seen in Fig.\ \ref{sel.plot}
these values suggest that the median galaxy is an
$L\approx 1.2 L_*$ early-type galaxy at redshift $z\approx 0.11$.
Currently we have redshifts for only nine of the 123 galaxies:
four in the CDF-S from COMBO-17 (Wolf et al.\ 2005), two in the
SDSS\,1030+05 field from the SDSS ({York} {et~al.} 2000), and three in the
CW\,1255+01 field from the SDSS.
The mean redshift of these nine objects
$\langle z \rangle = 0.13$ with a spread of 0.02,
consistent with the expectations from our color selection.

As a further test on the selection we obtained photometric and
spectroscopic data in ten randomly selected $3^{\circ}
\times 3^{\circ}$ fields from the SDSS. Galaxies were selected in
the $g-r$ vs.\ $r$ plane in the same manner as in our study,
using the Fukugita et al.\ (1995) transformations to convert
our $R$ and $B-R$ limits to SDSS $r$ and $g-r$ limits.
The median redshift is 0.098, with 90\,\% of the
galaxies at $z>0.05$. The rms field-to-field variation
of the median is only 0.008.

In the following we will adopt $z=0.1$ as the median redshift; our
conclusions change very little if we were to use, e.g., $z=0.08$ or
$z=0.13$ instead. For
$z\approx 0.1$ our observed $R$-band limit corresponds to $M_R \sim
-21$, and the median galaxy has $M_R \sim -22$.

\begin{figure*}[t]
\epsfxsize=18.3cm
\epsffile[14 14 596 596]{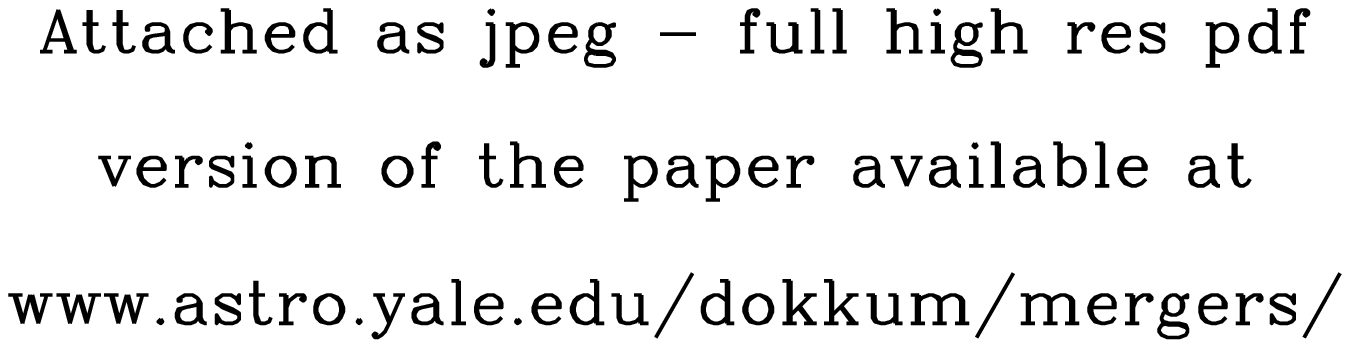}
\caption{\small
Examples of red mergers, ordered by the progression of the
interaction. The images were generated by combining the $B$ and $R$
frames. The objects are 17-596 and 17-681 (a); 19-2206 and 19-2242
(b); 1256-5723 (c); and 16-1302 (d).
Panel (a) spans $5'\times 5'$; panels (b), (c),
and (d) span $2\farcm 5 \times 2\farcm 5$.
The tidal features are faint and red, and generally barely
visible in $B$. 
Similar features are
seen in a large fraction of our sample of 123
red galaxies, in particular among the bulge-dominated
early-type galaxies.
Images of all objects are given in the Appendix.
\label{example_col.plot}}
\end{figure*}

\begin{figure*}[t]
\epsfxsize=18.3cm
\epsffile[14 14 597 597]{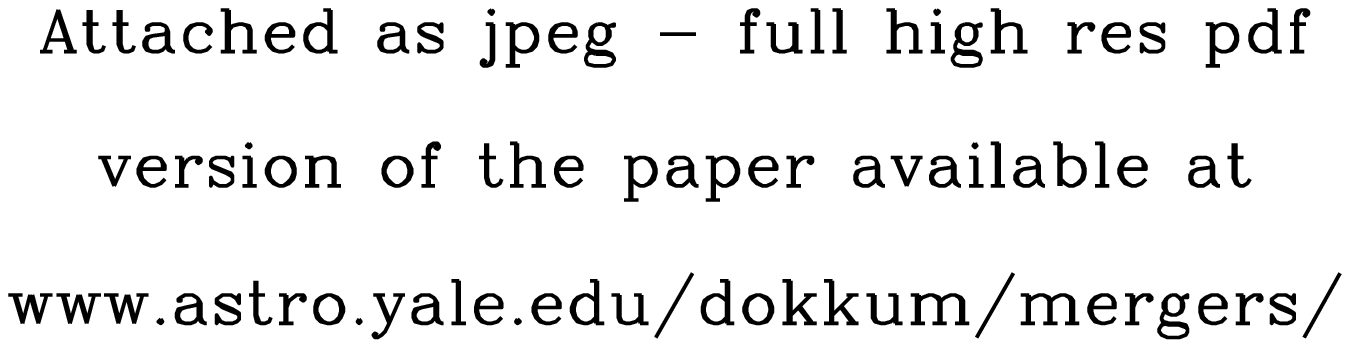}
\caption{\small
Same as Fig.\ \ref{example_col.plot}, but now highlighting faint
surface brightness levels in the summed exposures. The four
systems have extensive tidal debris, extending to $1'$
or more from the center. Features can be reliably detected
down to $\sim 28$\,\sbunit. Despite their large extent the
features contain less than 10\,\% of the total light of
the galaxies.
\label{example_bw.plot}}
\end{figure*}

\section{Analysis}

\subsection{Morphologies}
\label{class.sec}

All galaxies were assigned a morphological type by visually
inspecting their summed images (i.e.,
the ``BVR'' images for MUSYC galaxies and the ``BRI'' images
for NDWFS objects).
The images span a large range in
surface brightness levels, going
from the nearly saturated central regions to
very low surface brightness features up to $>50$\,kpc away from
the center. As the relevant dynamic brightness range  is $\gtrsim 10^4$
each galaxy was displayed
at four different contrast levels simultaneously.

The morphological types are
necessarily broad. Although the S/N ratio is very high
the spatial resolution is quite poor: the typical seeing is $1\farcs 1$,
which corresponds to 3\,kpc at $z=0.1$. Therefore, we only have
$2-3$ resolution elements within the half-light radii of many
galaxies (see, e.g., J\o{}rgensen, {Franx}, \& 
{Kj\ae{}rgaard} 1995). The assigned types
are spiral (S), indicating the presence of spiral arms and/or
star forming regions in a disk; S0, indicating an early-type
galaxy with an unambiguous disk component; and E/S0, indicating
a bulge-dominated early-type galaxy.
We cannot securely separate elliptical galaxies from bulge-dominated
or face-on S0 galaxies. Contrary to the usual definition
the E/S0 class therefore encompasses
ellipticals, bulge-dominated S0s, E/S0s, and S0/Es.

Morphological classifications are listed in Table 1, and images
of all 126 galaxies are shown in the Appendix. As expected, the
sample is dominated by early-type galaxies: of 126 objects, 10
(8\,\%) are classified as spirals, 30 (24\,\%)
as S0s (i.e., disk-dominated early-type
galaxies), and 86 (68\,\%) as E/S0s (i.e., bulge-dominated
early-type galaxies). Most of the
spiral galaxies have large red bulges and faint blue arms. Based
on their morphologies at surface brightness levels
$\mu \lesssim 25$\,mag\,arcsec$^2$ we infer that virtually all
galaxies in our sample are red because of their
evolved stellar populations, and not because of dust
(as is well known from many previous studies of bright
red galaxies in the local Universe; see, e.g.,
Sandage \& Visvanathan 1978, Strateva et al.\ 2001).

\subsection{Tidal Features}
\label{tidalclass.sec}

Tidal features were first identified by visual inspection of the full
sample of 126 galaxies; a quantitative
analysis of disturbances in the restricted sample
of 86 bulge-dominated early-type galaxies follows
in \S\ \ref{quant.sec}.
The flag describing tidal features can have
one of four values: 0 for no tidal features, 1 for weak
features, 2 for strong features, and 3 for an ongoing interaction with
another galaxy. Galaxies in the ``2'' class are generally
highly deformed merger remnants, whereas the ``1'' class indicates
more subtle features.  The difference is obviously quite
subjective, and in the subsequent analysis these two classes will
generally be combined into one.  A ``3'' classification implies that
both the primary and secondary galaxy show clear
distortions or tidal tails.

\begin{figure*}[b]
\epsfxsize=17.5cm
\epsffile[-105 14 659 558]{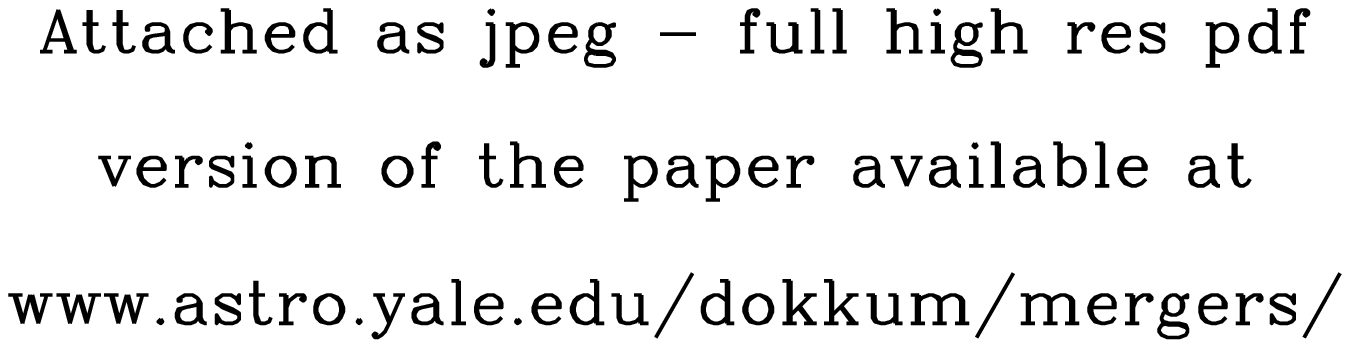}
\caption{\small
Illustration of the derivation of $t$, the quantity describing the
deviations from ellipse fits to the galaxies. Three galaxies are
shown: 7-1818 (top) has no visually detected features; 18-794
(center) has a weak tidal feature, and 1256-5723 (bottom) has strong
tidal features. From left to right are shown: the summed galaxy
image; the galaxy image $G$ with sharp features, blue features, and
low S/N regions masked; the elliptical model fit $M$; and the
median-filtered, noise-corrected distortion
image $F$ from which $t$ is measured.
\label{tidalpix.plot}}
\end{figure*}

The key result of our analysis is the ubiquity
of tidal features, particularly among bulge-dominated
early-type galaxies.
In the full sample of 126 galaxies, 44 (35\,\%) show clear signs of past
interactions and in an additional 23 cases (18\,\%) the interaction
is still in progress. Only 59 galaxies (47\,\%) appear undisturbed,
showing no unambiguous tidal features
at the surface brightness limit of the survey.
The fraction of disturbed objects is lowest among galaxies with
a clear disk component. Among 40 galaxies classified as
S or S0, only six (15\,\%) show evidence for past or present
interactions. In contrast, among the 86 galaxies classified as E/S0
undisturbed objects are the exception,
as 61 (71\,\%) show tidal features.

The nature and extent of the disturbances span a wide range.
Some galaxies have clearly defined tidal tails while others
show broad fans
of stars similar to the final frames of the simulation shown
in Fig.\ \ref{sim.plot}. In most cases the disturbances are
subtle and only visible at large radii and
very faint surface brightness levels,
although some objects are strongly disturbed throughout.
Often we see a mixture, e.g., a well defined tail in addition
to a broad, smooth disturbance at very faint levels.
This large range of properties is not surprising as it
reflects the variation in the age of the interaction,
the viewing angles, and the properties of
the progenitors.

The tidal features are
almost always red. In many cases the features are only visible in the
$R$ and $I$ frames
despite the substantial depth of the blue exposures.
They also appear smooth, showing
no or very little evidence for clumps and
condensations. In these respects the features are
very different from the blue, clumpy tidal tails seen in spiral--spiral
interactions (e.g., {Mirabel},
{Dottori}, \& {Lutz} 1992; {Hunsberger}, {Charlton}, \&  {Zaritsky} 1996),
and from the sharp shells and ripples detected in unsharp-masked images
of ellipticals (e.g., Schweizer \& Seitzer 1992, Colbert et al.\ 2001).
We note here that our data are not well suited to identify
narrow features, as the FWHM resolution of our data is about 2 kpc
at $z=0.1$.

The remarkable nature of these red mergers and their remnants
is illustrated in Fig.\ \ref{example_col.plot}. The
figure shows
two examples of ongoing mergers, an example of a strongly
disturbed merger remnant, and an example of a galaxy with
more subtle distortions at faint surface brightness levels,
arranged in a plausible red merger sequence. In all cases
the tidal features are smooth and red, and quite different
from the highly structured blue tails associated
with known mergers between gas rich disk galaxies.
Figure \ref{example_bw.plot}
shows summed images of the same objects, highlighting the
faintest features and providing an indication of the
surface brightness levels reached by the observations.
We stress that these objects are fairly typical examples:
as can be inferred from
the smaller images shown in the Appendix they are by no
means unique within our sample.

\subsection{Quantitative Identification of Tidal Features}
\label{quant.sec}

Quantitative characterization of the tidal features is important for
testing the robustness of the visual classifications,
measuring the flux associated with the
features, assessing the
effects of changing the S/N ratio,
and examining correlations between distortions and other
properties of the galaxies. Quantitative criteria are also
useful as a tool for future studies of larger samples.
Standard measures of asymmetry (e.g.,
Abraham et al.\ 1996; Conselice et al.\ 2003) are not applicable to
these galaxies as the features comprise only a small percentage of the
total luminosity of the galaxies. Instead a method was
developed which determines distortions with respect to a
model light distribution (see, e.g., Colbert et al.\ 2001).
The method is only meaningful for bulge-dominated early-type
galaxies, which make up the bulk of the sample.

First, the galaxies are fitted by an elliptical galaxy model using
the ``ellipse'' task in IRAF, in three iterations. After
each iteration a mask file is updated using the residuals from
the previous fit. The center position, ellipticity,
and position angle
are allowed to vary with radius. In cases where a second galaxy
overlaps the primary object the two objects are fitted
iteratively. The final model image is denoted $M$.
Next, a ``clean'' galaxy image $G$ is produced in the following
way. Objects bluer or redder than the primary
galaxy are masked, by dividing the $R$ and $B$ band
images, median filtering, and identifying pixels deviating more than
a factor two from the median color of the galaxy. To remove
small foreground and background objects and retain the smooth
galaxy light a ``reverse'' unsharp masking technique is used:
the galaxy images are compared to Gaussian-smoothed versions
of themselves, and pixels deviating more than a factor two
are masked (excluding the galaxy centers).
Pixels in the vicinity of masked pixels are also masked.
Finally, a fractional
distortion image $F$ is created by dividing $G$ by $M$.
The distortion image is convolved with a $5 \times 5$ median
filter to reduce pixel-to-pixel variations.
The parameter $t$ describing the level of distortion
is defined as
\begin{equation}
t = \overline{\left| F_{x,y} - \overline{ F_{x,y} } \right| },
\end{equation}
The tidal parameter thus measures the median absolute
deviation of the (fractional) residuals from the
model fit.

The procedure is illustrated in Fig.\ \ref{tidalpix.plot},
for a galaxy pair with no visible distortions, a galaxy with
a weak tidal feature, and a strongly disturbed object.
The visually identified tidal features are isolated
and emphasized in the distortion images $F$ shown at
right. The corresponding values of $t$ vary from 8\,\%
for the undistorted object to 24\,\% for the strongly disturbed
object.

In Fig.\ \ref{comp.plot}(a) the values of $t$ are compared to
the visual tidal classifications for all 86 bulge-dominated early-type
galaxies. There is a strong correlation:
the median value of $t$ is 0.08 for galaxies classified as
undisturbed, 0.13 for weakly disturbed galaxies, and 0.19
for strongly disturbed galaxies and ongoing mergers.
Tidal features are usually visually identified if $t>0.1$:
of 23 galaxies with $t<0.1$ only 5 (22\,\%) were visually classified
as tidally distorted, compared to 56 out of 63 galaxies
with $t\geq 0.1$ (89\,\%). We conclude that the visually
identified distortions are robust and imply median
absolute deviations from an ellipse fit of $\gtrsim 10$\,\%.

\vbox{
\begin{center}
\leavevmode
\hbox{%
\epsfxsize=8.5cm
\epsffile{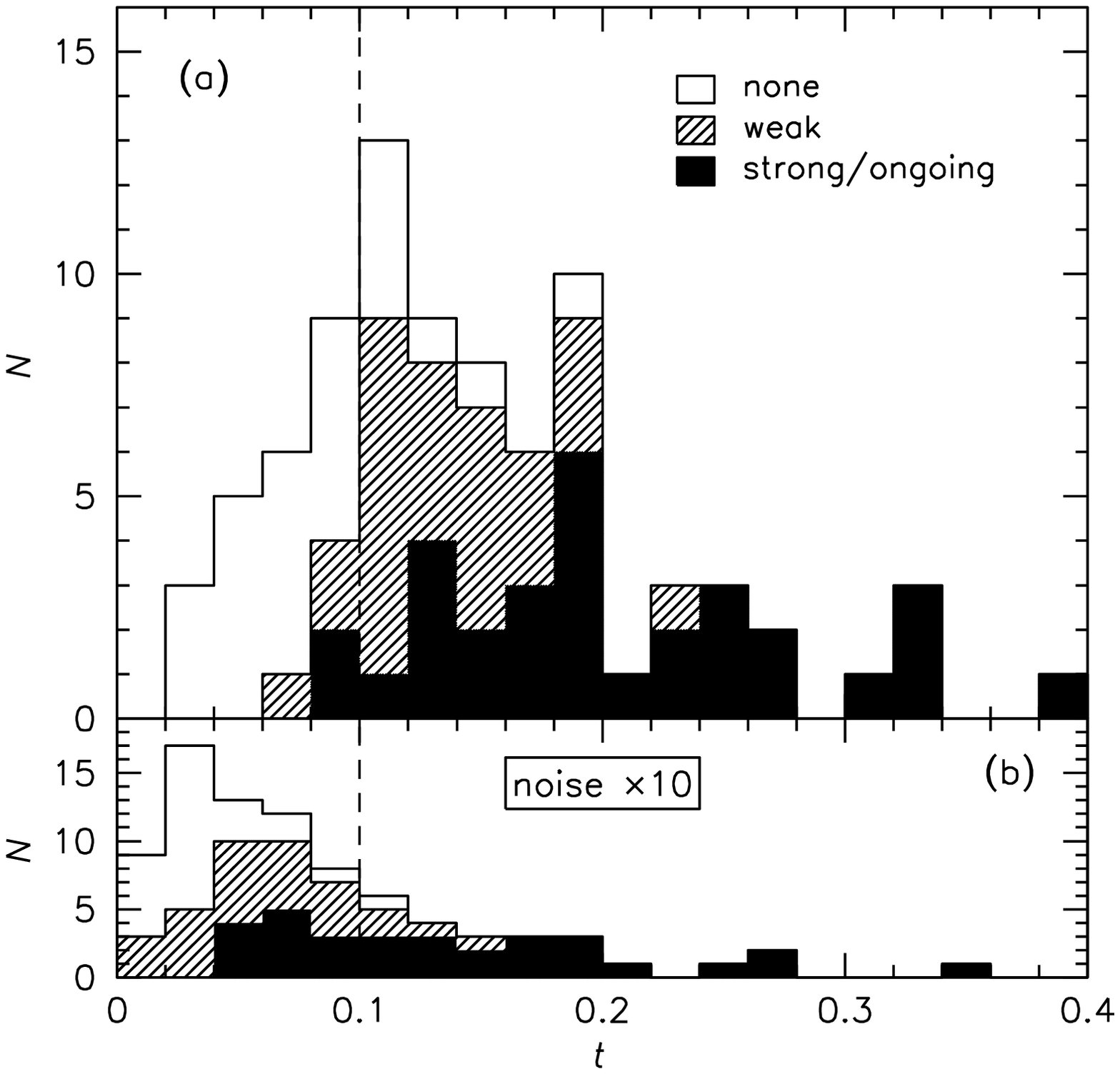}}
\figcaption{\small
Comparison of visual classifications and the tidal parameter $t$
for the 86 galaxies classified as E/S0. There is a strong
correlation between $t$ and the visual classifications.
Galaxies with visually identified tidal features generally
have $t>0.1$. The histograms in (b) show the effect of
lowering the S/N ratio by a factor ten. The majority of
galaxies move to $t<0.1$, and hence would not be identified
as tidally distorted objects.
\label{comp.plot}}
\end{center}}

The distortion maps $F$ can be used to estimate the amount of
star light associated with the tidal features. Visual
inspection of the distortion maps suggests that pixels
deviating more than 15\,\% from the model are usually
associated with visible features. A mask $F'$
was created by setting pixels with values $\geq 0.15$ in $F$ to
1 and all other pixels to zero. The relative flux in the tidal features
was then estimated as follows:
\begin{equation}
f_t = \frac{\displaystyle\sum  F'_{x,y} \times (G_{x,y} -
M_{x,y})}{\displaystyle \sum M_{x,y}},
\end{equation}
with the summations over all pixels $x,y$.
As expected, the median value of $f_t$ is low for the 25 E/S0
galaxies with $t<0.1$: 0.01, about 1\,\% of the galaxy light.
The median is 0.04 for the 63 galaxies with $t\geq 0.1$
and 0.07 for the 14 most disturbed galaxies with $t\geq 0.2$.
From varying the cutoff in $F$ the systematic uncertainty
in these values is estimated at $\sim 30$\,\%. We infer
that -- despite their large extent --
the tidal features typically contain only about $5$\,\%
of the total light of the galaxies. 

\begin{figure*}[p]
\epsfxsize=18cm
\epsffile[14 14 594 739]{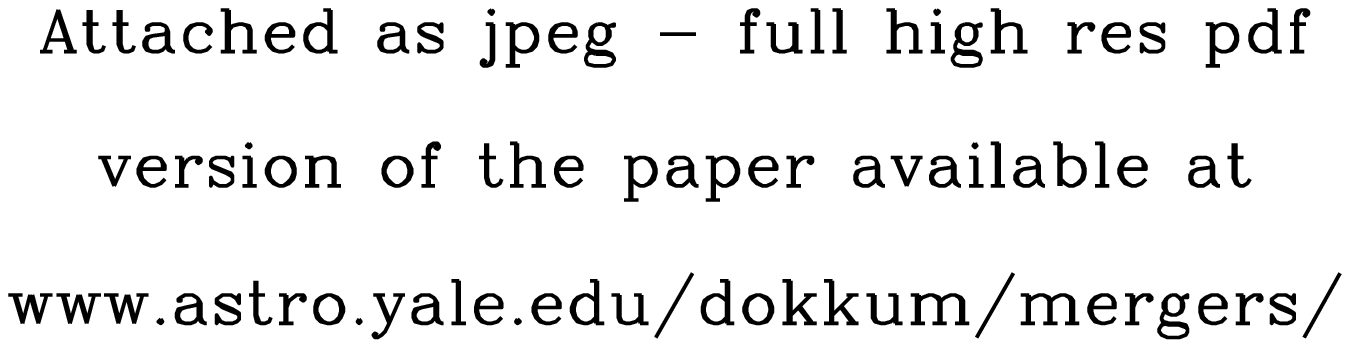}
\caption{\small
The nineteen ongoing mergers in the red galaxy sample.
The center of 11-962 is unresolved
at the resolution of the NDWFS, and the image showing the two nuclei
was obtained with the WIYN telescope in $0\farcs 45$ seeing.
Note that in four
cases both the primary galaxy and its companion are in the $R<17$
red galaxy sample. In three cases
(4-1190, 4-1975, and 22-2252) the individual paired galaxies
are fainter than $R=17$ but their combined luminosity
exceeds this limit. 
\label{ongoing.plot}}
\end{figure*}

\begin{figure*}[t]
\epsfxsize=17.5cm
\epsffile[32 175 543 425]{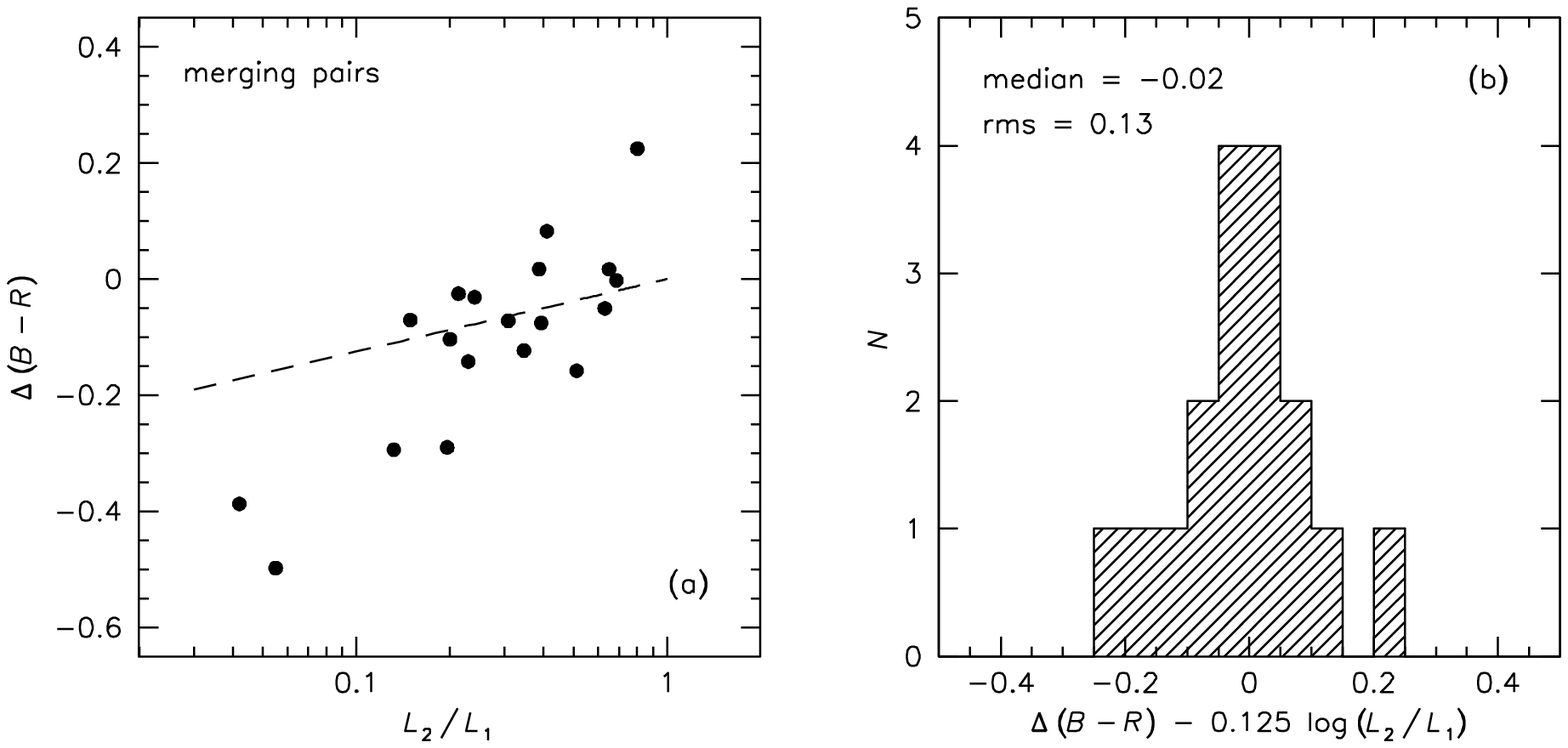}
\caption{\small
Luminosity ratios and color differences of the 19 interacting pairs.
The median luminosity ratio is 0.31, and the median color difference
is $-0.07$. Panel (a) shows that the color difference correlates with
the luminosity ratio, in the sense that faint companions are bluer.
The broken line is the correlation expected for
early-type galaxies on the color-magnitude relation. Color differences
are plotted in (b), after removing this relation. The median residual
color difference is $-0.02$, with a scatter of $0.13$ magnitudes.
\label{ratio.plot}}
\end{figure*}

\subsection{Ongoing Mergers}
\label{ongoing.sec}

There are 23 galaxies in the sample that show
a clear tidal connection to a secondary object. In four
cases the secondary object is also in the sample of 126,
leaving nineteen unique systems. The fact that tidal
features have developed implies that dynamical friction
is already at work. Assuming that the galaxies were initially
on nearly parabolic orbits, which is a reasonable assumption
for field galaxies, the implication is that virtually all these
galaxies will eventually merge
(J.\ Barnes, private communication).
We note that our
selection of merger pairs by tidal features is different from
``standard'' photometric and spectroscopic
selection of close pairs, where the number of mergers
is always smaller than the number of close pairs
(e.g., Patton et al.\ 2002): as expected, there are several close
pairs in our sample which do not show tidal features
and are not classified as ongoing mergers
(see, e.g., the top panels of Fig.\ \ref{tidalpix.plot}).


The 19 merger systems are shown in
Fig.\ \ref{ongoing.plot}; color images of these objects
are in the Appendix,
along with all other galaxies.
In most cases the primary galaxy is
in the sample of 126 and the secondary galaxy is not (because
it has $R>17$).
In cases where both galaxies are in the sample, and in
cases where the primary galaxy has a double nucleus, we
designated the brightest object or nucleus
as the primary object. The primary galaxies in the merging
pairs are equally bright as other
galaxies in the sample:
their median $R$ magnitude is 16.3,
compared to 16.4 for the full sample
of red galaxies.

The merger sample comprises objects connected by a tidal
``bridge'' (e.g., 1-2874, 5-2345, 18-485/522); double nuclei
in a highly disturbed common envelope (cdfs-374, cdfs-1100,
4-1190, 11-962); and objects of similar brightness with disturbed
isophotes and tidal tails or fans (e.g., 2-3070/3102,
4-1975, 7-4247, 17-596/681, 19-2206/2242).
Visual inspection of the merging systems indicates that the mergers
are usually not between blue, disk-dominated systems but between red,
bulge-dominated systems. Furthermore, in many cases the secondary
galaxy appears to be of similar brightness as the primary galaxy.
Eight of the 126 galaxies are merging with each other: the four
bright, red pairs are 2-3070/3102, 17-596/681, 18-485/522, and
19-2206/2242.

We quantified these effects in the following way. The luminosity
ratio is defined as
\begin{equation}
\frac{L_2}{L_1} = 10^{(R_1 - R_2)/2.5},
\end{equation}
and the color difference as
\begin{equation}
\Delta (B-R) = (B_2-R_2) - (B_1-R_1).
\end{equation}
In the 16 cases
where the primary and the secondary galaxy are both in the
SExtractor catalog we
calculated the luminosity ratio and the color
difference directly (see
\S\,\ref{cat.sec}).  In three cases the pair is too close to be
separated into two objects by SExtractor (cdfs-374, cdfs-1100, and
11-962). These objects are perhaps better described as single galaxies
with double nuclei.
Photometry for the nuclei was obtained from
aperture photometry, with the radius of the aperture equal to half the
distance between the nuclei.
We note that the double nucleus of 11-962 is
elongated but unresolved in the NDWFS images: the enlarged image shown
in Fig.\ \ref{ongoing.plot} was obtained with the OPTIC camera
({Tonry} {et~al.} 2005)
on the WIYN telescope, and has a resolution
of $0\farcs 45$ FWHM.

As a check on the robustness of the results we also obtained aperture
photometry in fixed $5\arcsec$ apertures for the 16 pairs that are
well separated. This method underestimates the luminosity differences,
as the secondary galaxies are usually more compact than the primary
galaxies. Nevertheless, the results are very similar, giving a median
luminosity ratio that is only $\sim 20$\,\% higher than derived from
the SExtractor AUTO magnitudes.

The pair photometry is given in Table~1.
Luminosity ratios range from 0.04 to 0.80, i.e., from relatively minor
1:25 accretion events to nearly equal mass mergers. The median luminosity
ratio is 0.31, or a 1:3 merger. We conclude
that approximately half the ongoing interactions are major mergers.
The median color difference is only
$-0.07$, i.e., companions are typically 0.07 magnitudes bluer than
the primary galaxy.
As the companions are (by definition) fainter than
the primary galaxies, this small difference may be an effect
of the existence of the color-magnitude relation.
Figure \ref{ratio.plot}(a) shows the relation
between color difference and luminosity ratio.
There is a correlation,
with the faintest companions being the bluest. The dashed line
shows the expected correlation for elliptical galaxies on the
(galaxy cluster) color-magnitude relation ({L{\' o}pez-Cruz}, {Barkhouse}, \&  {Yee} 2004).
In Fig.\ \ref{ratio.plot}(b) we show the distribution of
residual color differences
after subtracting this relation. The median of the distribution
is $-0.02$ and the rms scatter is only $0.13$ magnitudes.

These results show that the ongoing mergers in our sample occur ``within''
the red sequence with very small scatter. This result cannot be
attributed to our selection criteria: as 16 of the 19
pairs are separated by SExtractor there is no obvious bias against
detecting blue companions to the primary galaxies. There is a hint that
the observed relation in Fig.\ \ref{ratio.plot}(a) is steeper
than expected from the color-magnitude relation, but this is
largely due to the two pairs with $L_2/L_1 \approx 0.05$
and the spiral/S0 merger 18-485/522. Among pairs with $L_2/L_1>0.1$
whose primary galaxy is classified
as early-type the residual scatter
in $\Delta (B-R)$ is only 0.08 magnitudes.
\vspace{0.5cm}

\begin{small}
\begin{center}
{ {\sc TABLE 1}\\
\sc Photometry of Merger Pairs} \\
\vspace{0.1cm}
\begin{tabular}{lcr}
\hline
\hline
Object & $L_2/L_1$ & $\Delta (B-R)$ \\
\hline
cdfs-374 & 0.39 & $ -0.08 $ \\
cdfs-1100 & 0.35 & $ -0.12 $ \\
cdfs-6976 & 0.23 & $ -0.14 $ \\
1-2874 & 0.05 & $ -0.50 $ \\
2-3070 & 0.63 & $ -0.05 $ \\
4-1190 & 0.39 & $ 0.02 $ \\
4-1975 & 0.51 & $ -0.16 $ \\
4-2713 & 0.13 & $ -0.29 $ \\
5-2345 & 0.04 & $ -0.39 $ \\
7-4247 & 0.21 & $ -0.03 $ \\
11-962 & 0.69 & $ -0.00 $ \\
11-1278 & 0.15 & $ -0.07 $ \\
11-1732 & 0.20 & $ -0.29 $ \\
14-1401 & 0.24 & $ -0.03 $ \\
17-596 & 0.65 & $ 0.02 $ \\
18-485 & 0.80 & $ 0.22 $ \\
19-2206 & 0.41 & $ 0.08 $ \\
22-2252 & 0.31 & $ -0.07 $ \\
26-2558 & 0.20 & $ -0.10 $ \\
\hline
\end{tabular}
\end{center}
\end{small}

\section{Discussion}

\subsection{Why Were They Missed?}

The key results of our analysis are the large number of ongoing
red mergers and the ubiquity of tidal features associated with
(in particular) bulge-dominated early-type galaxies (galaxies
classified as ``E/S0''). As discussed in
\S\,\ref{tidalclass.sec} the large scale, low surface brightness
features that
we see are different from the ripples and shells that have
been reported previously in unsharp-masked images of ellipticals
at $z\sim 0$.
Our sample is essentially local, in the sense that the merger rate is
not expected to evolve significantly from $z\sim 0.1$ to $z\sim 0$. Nearby
galaxies have been studied in great detail over the past decades,
and before discussing the consequences of our findings 
we address the question why this preponderance of smooth
red tidal features has not been seen before.

The subjective nature of the visual classifications can be ruled
out as a cause given the consistency of visual and quantitative
classifications: 73\,\% of E/S0 galaxies have $t>0.1$.
Small number statistics
also play a minor role given the large number of objects
in the sample.
Given the fact that more than 90\,\% of the sample comes
from the NDWFS we cannot rule out that the
fraction of interacting galaxies is unusually high in that region
of the sky. Random $3^{\circ} \times 3^{\circ}$
SDSS fields (see \S\,\ref{z.sec})
show substantial peaks in the redshift distribution
within each
field, reflecting the fact that most red galaxies live in
groups and filaments. We note, however, that the red galaxies are distributed
rather uniformly over the NDWFS area, and that
some of the most spectacular mergers
are in the unrelated MUSYC fields. 

The most likely reason why the ubiquity of red mergers in the
local universe was missed so far is
the depth and uniformity of the available imaging. Classic
imaging studies of nearby elliptical galaxies
(e.g., {Franx}, {Illingworth}, \&  {Heckman} 1989; {Peletier} {et~al.} 1990) used exposure times in the red
of only 180\,s -- 600\,s  on 1\,m -- 2\,m
class telescopes, and such short exposures have been the norm
ever since (e.g., {Zepf}, {Whitmore}, \& {Levison} 1991; {Pildis}, {Bregman}, \&  {Schombert} 1995; {Jansen} {et~al.} 2000; Colbert et al.\ 2001).
Similarly, the effective exposure time of the Sloan Digital Sky Survey
is about 51\,s per filter ({York} {et~al.} 2000).

Among the deepest imaging surveys are those of {Malin} \& {Carter} (1983),
who performed extreme enhancements of photographic plates to bring
out low surface brightness features, and {Schweizer} \& {Seitzer} (1992),
who used 2400\,s -- 3600\,s integrations with the 0.9\,m telescope
on Kitt Peak.
These studies reach AB surface brightness limits of $\mu \sim 26.5$\,\sbunit.
Although both studies find sharp distortions in the form of ripples
and shells in a large fraction of ellipticals, their depth and
field-of-view are not sufficient to find the large low surface
brightness features that we report here.
The exposure time that went into each of the images discussed here
is $\approx 27,000$\,s on 4\,m class telescopes equipped
with modern CCDs (equivalent to 120 hours on a
1\,m class telescope), and to our knowledge
such long exposures have never been obtained of a significant sample
of nearby red galaxies. 

Figure \ref{bad.plot} shows what Fig.\ \ref{example_bw.plot}
would look like if we had exposed for 600\,s on a 1\,m class
telescope. Very little remains of the dramatic tidal
features evident in Fig.\ \ref{example_bw.plot}, illustrating
the extreme depth of the NDWFS and MUSYC surveys.
We quantified the effect of the S/N ratio on the detectability
of tidal features by artificially
increasing the noise in our images by a factor
of ten, corresponding to a decrease in exposure
time of a factor of 100. The degraded images are of similar depth
as those of
Malin \& Carter (1983) and Schweizer \& Seitzer (1992),
and are still substantially deeper than virtually all other
imaging studies of nearby elliptical galaxies.
As shown in
Fig.\ \ref{comp.plot}(b) the distribution of $t$ changes
substantially when the S/N is decreased.
Only 29\,\% of E/S0s in
the degraded images have $t>0.1$,
compared to 73\,\% in the original images.

We conclude that the high incidence of tidal features in our
sample is a direct consequence of the faint surface brightness
levels reached by the data.
We also note that flat fielding uncertainties
may prohibit the detection of broad tidal features
around local galaxies irrespective of exposure time.
The faint debris in Fig.\ \ref{example_bw.plot}
has a very large extent 
in comparison to the high surface brightness
regions visible in Fig.\ \ref{bad.plot}, and moving the galaxies to
$z\sim 0.01$ would change the surface brightness
very little but increase the sizes of the debris fields
to $\sim 10'$ or more.
Finally, galaxy surveys in blue filters
(e.g., {Arp} 1966) would classify the
majority of objects in our sample as undisturbed even if
they met the
surface brightness and flat fielding requirements.

\vbox{
\begin{center}
\leavevmode
\hbox{%
\epsfxsize=8.5cm
\epsffile{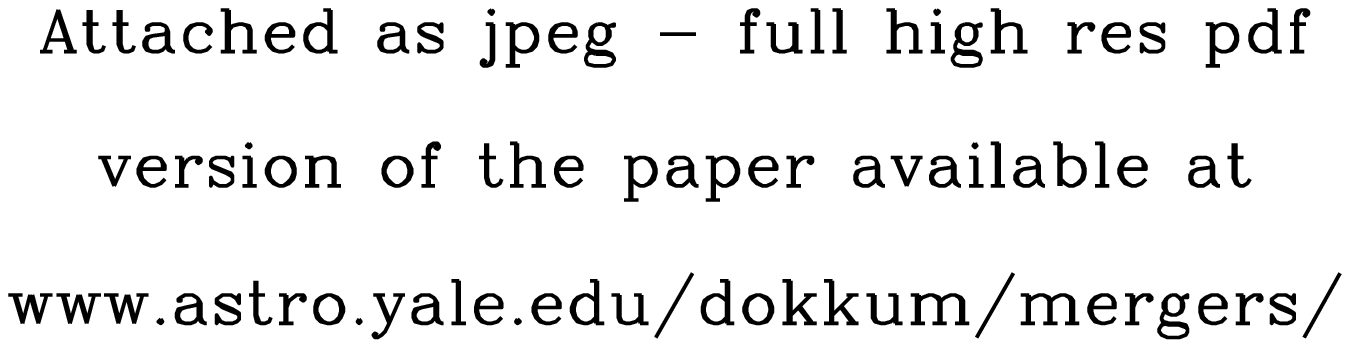}}
\figcaption{\small
Simulated appearance of the galaxies in
Figs.\ \ref{example_col.plot} and \ref{example_bw.plot} if
observed for 600\,s on a 1\,m class telescope,
a typical imaging depth for studies of nearby bright
galaxies. Owing to its
low surface brightness the large scale tidal debris
is all but invisible. For comparison,
the contour outlines the $\mu \approx 27.5$\,\sbunit\
level in Fig.\ \ref{example_bw.plot}. 
\label{bad.plot}}
\end{center}}

\subsection{Recent Merger History of Bulge-Dominated Galaxies}
\label{observedmerge.sec}

We first determine the implications of the observed interactions
only, without applying corrections for the short duration
of the mergers.
We define a sample which contains both
current and future bulge-dominated galaxies, consisting of
the 86 bulge-dominated
galaxies (classified as E/S0),
minus half of the six E/S0s which are interacting with each
other, plus the disk-dominated galaxies (classified as S
or S0) which are involved in a major merger.
Among disk-dominated galaxies the only major merger
is the spiral/S0 pair 18-485/522, whose constituent
galaxies have roughly equal luminosity.
The total sample of current and future
bulge-dominated galaxies is therefore $86 - 3 + 1 = 84$.
Among this sample of 84 galaxies there are eighteen current mergers and
41 remnants, and we infer that 70\,\% of current and future
bulge-dominated galaxies experienced a merger or accretion event in
the recent past.

The red ongoing mergers and the red tidal features associated
with many of the E/S0 galaxies
very likely sample the same physical process at different times.
On average, galaxies with strong tidal features are probably
observed shortly after the merger and galaxies with weak
features are observed at later times. Assuming that the
ongoing mergers are representative for the progenitors of all
remnants we can directly infer the mass ratios of the progenitors
of the full sample of 59 current and future bulge-dominated
early-type galaxies.
The median luminosity ratio of the ongoing mergers is 0.31, and
the median color difference is negligible after correcting for the
slope of the color-magnitude relation. Assuming that
$M/L \propto M^{0.2}$ (e.g., J\o{}rgensen et al.\ 1996)
a luminosity ratio of 0.31 implies a median mass ratio of 0.23,
or a 1:4 merger.

There is also indirect evidence
that the progenitors of
the remnants were typically major mergers rather than low mass
accretion events. Simulations by {Johnston}, {Sackett},
\&  {Bullock} (2001) show
that surface brightness levels $28<\mu<33$ are typical
for the debris of small satellites such as the Local Group
dwarfs.
Similarly, the average
surface brightness of the giant stream of M31 is
$\mu_V
\approx 30$\,\sbunit\ (Ibata et al.\ 2001) and that
of the debris from the Sagittarius dwarf
$\mu_V \approx 31$\,\sbunit\ (Johnston et al.\ 2001).
All these values are well beyond the detection limit of our
survey. Furthermore,
the fraction of tidally disturbed galaxies is much
lower among disk-dominated galaxies than among bulge-dominated
galaxies. Ignoring the ongoing mergers
only 8\,\% of disk-dominated galaxies show tidal
features compared to
62\,\% of bulge-dominated galaxies.
This difference is consistent with the idea that the events
responsible for the tidal
features were usually sufficiently strong to
disturb any dominant disk component.
Finally, the fact that the features are typically
broad and red suggests that the progenitors were dynamically
hot systems with old stellar populations, consistent with the
properties of the galaxies in the ongoing mergers.
We also note that the fraction of light associated with the features
($\sim 5$\,\%) is very similar to the fraction of light
associated with the tidal debris in the 1:3 merger simulation
shown in Fig.\ 1.

We conclude that approximately 35\,\% of
bulge-dominated red galaxies experienced a major
merger with mass ratio $>1:4$ in the
time window probed by our observations. This result
is direct observational confirmation of the
hierarchical assembly of massive galaxies.

\subsection{Merger Rate and Mass Accretion Rate}
\label{rate.sec}

Up to this point the analysis did not require
an estimate of the timescale
of the mergers. Such estimates are obviously uncertain, but
they are necessary for turning the merger fraction into a merger
rate and a mass accretion rate. These numbers
are more easily compared to models and other
observational studies,
and are needed for extrapolating the results to higher redshifts.

The merger rate can be defined in a variety of ways (see,
e.g., Patton et al.\ 2002). Here it is expressed as the number
of remnants that are formed per Gyr within our selection
area:
\begin{equation}
R = \frac{f_{\rm m}}{t_{\rm m}}\,\,{\rm Gyr}^{-1},
\end{equation}
with $f_{\rm m}$ the fraction of the galaxy population
involved in a merger --
with pairs counted as single objects -- and $t_{\rm m}$ a characteristic
timescale for the mergers.
We restrict the analysis to the sample of nineteen ongoing mergers,
as simulations of the fading of tidal debris around the remnants of
dry mergers have not yet been done in a systematic way.

Following Patton et al.\ (2000)
we assume that the timescale of the mergers can be approximated by
the dynamical friction timescale, given by
\begin{equation}
T_{\rm fric} = \frac{2.64 \times 10^5 r^2 v_c}{M \ln\,\Lambda}\,\,{\rm Gyr},
\end{equation}
where $r$ is the physical separation of the pairs, $v_c$
is the circular velocity, $M$ is the mass of the lowest
mass galaxy, and $\ln\,\Lambda$ is the Coulomb logarithm
(see Binney \& Tremaine 1987; Patton et al.\ 2000).
The median projected separation of the paired galaxies is
$8\farcs 6$, or $16\pm 3$ kpc
at $z=0.10 \pm 0.02$. Assuming random orientations
this corresponds to a median physical separation $r=20\pm 4$ kpc.
To obtain an estimate of $v_c$ we assume that the pairs have
similar line-of-sight velocity differences as the seven
elliptical-elliptical
mergers discussed in Combes et al.\ (1995). The mean velocity
difference of the Combes et al.\ pairs $\Delta v = 296 \pm 91$\,\kms
(with the error determined by the jackknife method), implying
$v_c = \sqrt{3} \Delta v = 513 \pm 158$\,\kms\ for an isotropic
velocity distribution. The median mass of the companions is
$(7 \pm 3) \times 10^{10}\,M_{\odot}$, where we used $M/L_R = 4.6$ in
Solar units to convert luminosity to mass (van der Marel 1991)
and assumed $z=0.10\pm 0.02$.
Taking $\ln \Lambda \sim 2$ (following Dubinski et al.\ 1999
and Patton et al.\ 2000) we obtain $T_{\rm fric} = 0.4 \pm 0.2$\,Gyr.
With $f_m = 19/122 = 0.16 \pm 0.03$ we obtain $R = 0.4 \pm 0.2$\,Gyr$^{-1}$.

\begin{figure*}[b]
\epsfxsize=17.5cm
\epsffile[44 486 517 654]{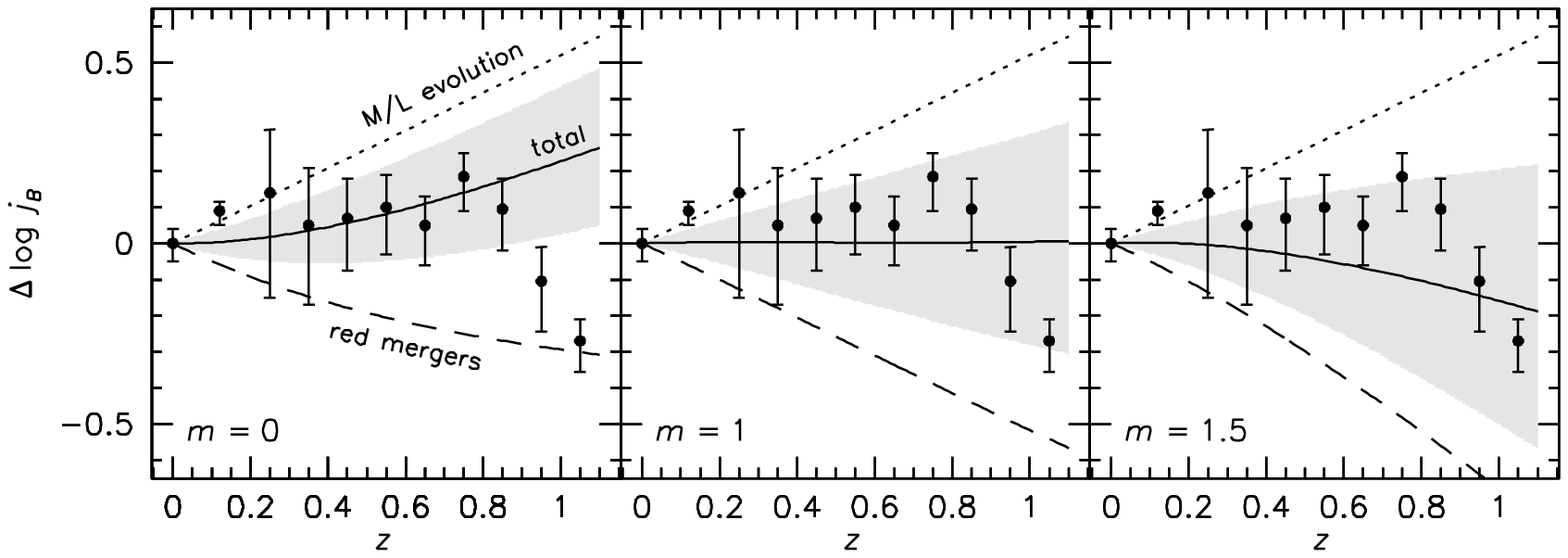}
\caption{\small
Evolution of the $B$-band luminosity density of red galaxies.
The dotted lines show the
evolution of the $M/L_B$ ratio of field
early-type galaxies,
as measured by van der Wel et al.\ (2005).
Dashed lines show the counter-acting effect of red mergers,
with different assumptions for the evolution of the mass accretion rate.
The mass accretion rate is $9\,\% \pm 4\,\%$ per Gyr at $z=0.1$, and evolves
as $(1+z)^m$. The solid line shows the
total luminosity density. Grey bands show the combined uncertainty
resulting from the uncertainties
in the merger rate and the evolution of the $M/L_B$ ratio.
Data points are from the Bell et al.\ (2004) analysis of the
COMBO-17 dataset. There is good agreement between the models
and the data, in particular if $m>0$.
\label{jevo.plot}}
\end{figure*}

The effect of the mergers on the mass evolution of red galaxies
not only depends on the merger rate but also on the mass change
resulting from individual mergers.
The mass accretion rate can be approximated by
\begin{equation}
\Delta M/M =  R \times \overline{M_2/M_1}\,\,{\rm Gyr}^{-1},
\end{equation}
with  $\overline{M_2/M_1}$
the median mass ratio of the mergers. As shown in
\S\,\ref{observedmerge.sec} this ratio is approximately
0.23. With
$R_{\rm m} = 0.4$ we obtain $\Delta M/M = 0.09
\pm 0.04$\,Gyr$^{-1}$, i.e., merging increases
the masses of galaxies
on the red sequence by $\sim 10$\,\% every $10^9$ years.

For comparison to other studies it is also of interest to consider
the major merger fraction within a projected separation of 20\,kpc.
There are seven red pairs with luminosity ratio $>0.3$ and projected
separation $<11\arcsec$, corresponding to a fraction of $0.06\pm 0.02$.
It should be stressed that this number refers to
mergers within the red sequence, not to the merger fraction
within the full sample of $R<17$ galaxies. The
colors of the well-separated
pairs show that red galaxies ``prefer'' to merge with other
red galaxies. We have not examined the prevalence of
mergers among luminous blue galaxies, but as discussed in
\S\,1 the stellar populations
of ellipticals rule out widespread major mergers among
this population.
Therefore, the major merger rate in the full sample of red
and blue galaxies is presumably much lower than that within the
restricted sample of red galaxies. A very rough estimate of the
merger fraction in the full sample is $0.06 \times
N_{\rm red}/N_{\rm total} \sim 0.02$, in reasonable
agreement with previous studies of close pairs
(see, e.g., {Patton} {et~al.} 2002, Lin et al.\ 2004, and
references therein).

\subsection{Effects on the Evolution of the Luminosity Density}
\label{lumfunc.sec}

At redshifts $z<1$, the observed evolution of the luminosity function of
red galaxies reflects passive evolution of the stellar
populations and possible changes in the underlying mass function.
As discussed in, e.g., McIntosh et al.\ (2005)
these changes can be due to mergers, galaxies entering
the red sample due to changes in their star formation rate, or other
effects.
The best available constraints on the evolution of the luminosity function
of red galaxies were derived by {Bell} {et~al.} (2004), using the
COMBO-17 survey. They find that the luminosity density of luminous red
galaxies is approximately constant out to $z\sim 1$, which is
surprising given the expected evolution of a factor of $3-4$
in the $M/L$ ratios of the galaxies
(e.g., {van Dokkum} {et~al.} 1998a,
Treu et al.\ 2005, van der Wel et al.\ 2005).
A possible explanation is that
the underlying stellar mass density evolves
as well, compensating for passive evolution of the stellar populations
({Bell} {et~al.} 2004).

We first determine the effect of the observed red mergers only, i.e.,
the 52\,\% of red galaxies that are merger remnants or a merger
pair. Dry mergers have
no effect on the total luminosity density, but they have a strong
effect on the luminosity density of galaxies brighter than
a fixed magnitude.
The effect of a single generation of
mergers  can be approximated by
\begin{equation}
j(z) \approx \left( 1 - f_{\rm m} \left\langle \frac{L_2}{L_1 + L_2}
\right\rangle \right) j(0),
\end{equation}
with $j(z)$ the luminosity density of luminous
galaxies before the mergers and $j(0)$
the luminosity density after the mergers.
For $f_{\rm m}=0.52$ and
$\langle L_2 / L_1 \rangle = 0.3$ we find $j(z) \approx 0.88
j(0)$. We use Monte Carlo simulations to test this approximation
for a fiducial luminosity function with $\alpha = -0.6$
and $M_* = -19.9$ ({Bell} {et~al.} 2004). The luminosity function
is evolved backward in time by breaking 50\,\% of the
galaxies into pieces with luminosity ratio 0.3.
Calculating the
luminosity density for $M<-19$ 
gives $j(z) = 0.89 j(0)$, in very good agreement with the simple
estimate given above. The conclusion is that the effect of
the observed mergers and remnants on
the luminosity density of bright galaxies is small,
of order 10\,\%.

The observed interactions only
probe a relatively short period of at most a few Gyr.
In order to extrapolate the effect of the mergers
back in time 
we assume the following: 1) the mass
accretion rate at $z=0.1$ is $\Delta M/M= 0.09 \pm 0.04$\,Gyr$^{-1}$
(see \S\,\ref{rate.sec});
2) the change in luminosity density due
to mergers is proportional to the accreted luminosity; 3)
all red galaxies  are equally likely to undergo mergers;
and 4) the mass accretion rate
evolves as $(1+z)^m$.
The value of $m$ is treated as a free parameter:
observational constraints on the evolution of the merger rate
may not be quite consistent
(see, e.g., Patton et al.\ 2002, Concelice et al.\ 2003, Lin et al.\
2004), and
no studies have specifically considered the evolution of the pair
fraction among galaxies on the red sequence.

The dashed lines in
Fig.\ \ref{jevo.plot} show the predicted merger-driven
evolution of the luminosity density of bright red galaxies
with these assumptions,
for three values of $m$. The model with $m=0$ has
a constant accretion rate, and in the model with $m=1.5$ the
accretion rate is $3\times$ higher at $z=1$ than it is
at $z=0$. 
The dotted lines
show the evolution of the $M/L_B$ ratio of field
early-type galaxies, as measured by van der Wel et al.\ (2005).
These authors find $\Delta \ln M/L_B = (-1.20 \pm 0.18) z$
for galaxies with $M>2\times 10^{11}\,M_{\odot}$ (appropriate for
our sample), which is
consistent with the independent measurement by Treu et al.\ (2005).
The solid line shows the predicted evolution of $j_B$ when both mergers
and $M/L$ evolution are taken into account. Grey bands
indicate the combined uncertainties in
$\Delta \ln M/L_B$ and the merger rate.
The predicted evolution of $j_B$ depends rather strongly on the
assumed evolution of the mass accretion rate:
it is positive for a non-evolving
accretion rate, constant for $m=1$, and negative for $m>1$.

Solid points show the luminosity density in luminous red galaxies
as measured by {Bell} {et~al.} (2004) (their Fig.\ 5).
As discussed extensively by these authors the data are
inconsistent with passive evolution alone (dotted curves).
However, as can be seen in Fig.\ \ref{jevo.plot}
the data are fully consistent with models that include
the mass accretion rate measured here. The uncertainties in the
datapoints, the measured $M/L$ evolution, and the merger rate
are too large to distinguish between models with different
values of $m$, although the data at $z>0.8$ seem to favor
models with $m>0$. 

We infer that the cumulative effects of dry
mergers may be the dominant cause of the constant
luminosity density of luminous red sequence galaxies in the
COMBO-17 survey. There are several caveats: both the datapoints
in Fig.\ \ref{jevo.plot} and the $z=0.1$ mass accretion rate
have substantial errors, leading to a wide range of allowed
models; although the most recent generation of mergers appears
to be (nearly) dissipationless, this may no longer hold
at $z\gtrsim 0.5$; and the fact that disk-dominated red
galaxies probably evolve differently from bulge-dominated
ones is ignored.
The lack of evolution in the luminosity density of
bright red galaxies, if confirmed, is
probably due to a combination of effects with dry merging
a significant, but not the only, contributor
(see also McIntosh et al.\ 2005).

\section{Conclusions}


From an analysis of tidal features associated with bright red
galaxies
we find that $\sim 70$\,\% of bulge-dominated galaxies experienced a
merger with median mass ratio 1:4 in the recent past.
Expressed
in other ways, $\sim 35$\,\% of bulge-dominated galaxies experienced a
major merger involving $>20$\,\% of its final mass, and the
current mass accretion rate of galaxies on the red sequence
$\Delta M / M = 0.09 \pm 0.04$\,Gyr$^{-1}$.
Assuming a constant or increasing mass accretion rate with
redshift it is inferred that the stellar mass density
in luminous red galaxies has increased
by a factor of $\gtrsim 2$ over the redshift range $0<z<1$.

Neither of the two standard paradigms for elliptical
galaxy formation appears to be consistent with our
results: ``monolithic'' assembly
at high redshift or late assembly via mergers of gas-rich disk
galaxies. Instead, elliptical galaxies appear to have been
assembled in mergers of bulge-dominated,
red galaxies.
This ``dry'' form of merging is qualitatively consistent with
the high central densities of ellipticals
(Gao et al.\ 2004), their red colors and
uniform properties, the existence of the $M_{\bullet} - \sigma$ relation
(Wyithe \& Loeb 2005),
and their specific frequency of globular clusters.

It remains to be seen whether widespread dry mergers are
consistent with the slope and scatter of the color-magnitude relation
(see, e.g., Bower et al.\ 1998, Ciotti \& van Albada 2001).
The median merger
in our sample makes the brightest galaxy more luminous by $\sim 30$\,\%
and bluer by only $\sim 0.02$ mag in $B-R$, and its remnant will
lie within $\sim 0.03$ mag of the color-magnitude relation. However,
it is doubtful whether multiple generations of such mergers
can be accommodated, unless there is a strong correlation between
the masses of the progenitors (see Peebles 2002). 
It will also be interesting to see whether the evolution is consistent with
the number density of
massive galaxies at high redshift. Again,
the {\em observed}
mergers only have a $\sim 10$\,\% effect on the mass function
of red galaxies,
and the extrapolation to higher redshifts is obviously
still very uncertain. Furthermore, the evolution of the mass
function of all galaxies may be different from that of the subset of
galaxies on the red sequence.

The high merger rate confirms
predictions from hierarchical galaxy formation models
in a $\Lambda$CDM universe (e.g., {Kauffmann} {et~al.} 1993; {Kauffmann} 1996; {Cole} {et~al.} 2000; {Somerville}, {Primack}, \&  {Faber} 2001; {Murali} {et~al.} 2002).
Furthermore, semi-analytical models have predicted
that gas-poor mergers between
bulge-dominated systems, rather than mergers of disk systems,
are responsible for the formation of the most massive
ellipticals (Kauffmann \& Haehnelt 2000; and, in particular,
{Khochfar} \& {Burkert} 2003).
Quantitative comparisons of mass growth
are difficult as current models do not naturally produce red
field galaxies without star formation: the colors of field
ellipticals in the simulations
are too blue and their $M/L$ ratios too low
(e.g., {Kauffmann} 1996; {van Dokkum} {et~al.} 2001a).
Additional mechanisms, such as heating by active nuclei, appear to be required
to halt gas cooling and star formation in massive galaxies
(e.g., {Binney} 2004, Somerville 2004, Dekel \& Birnboim 2005,
Keres et al.\ 2005).
Our results imply that revised models which address
these issues should not only reproduce
the ``red and dead'' nature of ellipticals today but also of
their immediate progenitors -- which may occur naturally if the
progenitors have masses greater than some critical mass
(see Cooray \& Milosavljevi\'c 2005).


The main uncertainty in the observed merger
fraction is the possibility
that the Northern NDWFS field, which contains over 90\,\% of
the sample, is special in its frequency of tidally
disturbed objects. This seems unlikely
given its area of $\sim 400$\,Mpc$^2$ at $z=0.1$, but the issue of
field-to-field variations in the merger fraction will only be settled
conclusively when independent fields of similar size are studied
in the same way.
The main uncertainty in the merger {\em rate} and mass accretion rate
is the timescale of the mergers.
Although our estimates broadly agree with other studies
(see, e.g.,
Lin et al.\ 2004, and references therein), this may simply
reflect the fact that similar assumptions lead to similar results.
Modeling of red, gas-poor mergers has not been done in a systematic
way using modern techniques, and it will be interesting to see
what the timescales are for the initial coalescence
and the subsequent surface brightness
evolution of tidal debris. Specifically, modeling of the 19 merging
systems and 44 remnants presented here would provide much better
constraints on the merger rate and mass accretion rate, particularly
when more complete redshift information is available.

More detailed
observational studies of the mergers and their remnants may
help answer the
question {\em why} the mergers are red, i.e.,
what made the progenitors lose their gas?
If active nuclei prevent the cooling of
gas above some critical mass at early times
they may play the same role during mergers,
and it will be interesting to compare the degree of nuclear activity
in undisturbed galaxies, ongoing mergers, and remnants.
Also, sensitive diagnostics of young populations
(e.g., H$\delta$ line strengths and ultra-violet photometry)
can provide better constraints on the star formation histories
of the mergers and remnants. Finally, studies with higher
spatial resolution can provide information on the detailed
isophotal shapes (boxy or disky) of the remnants and their
progenitors, and on possible correlations between large scale
smooth distortions and the sharp ripples and shells that
have been reported in $z\approx 0$ ellipticals (see
Hernquist \& Spergel 1995).

Our study focuses on events that we can see today, and 
it will be very interesting to push the analysis to higher redshifts.
Although the most recent generation of
mergers could be largely
``dry'', previous generations likely involved blue galaxies
and/or were accompanied by strong star formation
(see, e.g., {Sanders} {et~al.} 1988). 
Unfortunately it will be difficult to identify the broad
red tidal features that we see here at significantly
higher redshift
due to the $(1+z)^4$ cosmological surface brightness dimming.
The limiting depth of the MUSYC and NDWFS images is $\sim 29$\,\sbunit\
($1\sigma$, AB). An
equivalent survey at $z\geq 1$ should cover an area
of $\gtrsim 1$ square degree and reach levels
of $\sim 31.5$\,\sbunit\
at $z=1$ and $\sim 33.5$\,\sbunit\
at $z=2$.
Even when (unfavorable) $K$-corrections are ignored these requirements
are well beyond the capabilities
of current ground- or space-based telescopes.
A more viable technique is to focus on the fraction of
red galaxies in pairs. Although pair statistics require large corrections
due to the short timescale of the mergers, pairs are
easily detectable out to high redshift with the Hubble Space Telescope
(see {van Dokkum} {et~al.} 1999). 
Based on the work presented here the merger fraction
among galaxies on the red sequence
should be $(0.06 \pm 0.02) \times (1+z)^m$ for separations $<20$\,kpc
and luminosity ratios $\geq 0.3$.

\acknowledgements{
This paper was made possible by the dedicated efforts of all the
individuals
behind the NOAO Deep Wide-Field Survey and the Multi-wavelength
Survey by Yale-Chile. Particular thanks go to Buell Januzzi
and Arjun Dey for initiating and executing the NDWFS and to
Eric Gawiser, who
is the driving force behind MUSYC. David Herrera is responsible
for reducing most of the MUSYC optical imaging data.
Marijn Franx and Jeff Kenney are thanked for useful discussions.
Comments of
the anonymous referee improved the paper
significantly. The NDWFS is supported by the National Optical
Astronomy Observatory (NOAO). NOAO is operated by AURA, Inc.,
under a cooperative agreement with the National Science
Foundation.
}

\bibliography{}

\begin{references}

\reference{} Abraham, R.~G., Tanvir, N.~R., Santiago, B.~X., Ellis, R.~S.,
 Glazebrook, K., \& van den Bergh, S. 1996, MNRAS, 279, L47

\reference{}{Arp}, H. 1966, \apjs, 14, 1

\reference{}{Balcells}, M. \& {Quinn}, P.~J. 1990, \apj, 361, 381

\reference{}{Barnes}, J. \& {Hut}, P. 1986, \nat, 324, 446

\reference{}{Becker}, R.~H., {Fan}, X., {White}, R.~L., {Strauss}, M.~A., {Narayanan},  V.~K., {Lupton}, R.~H., {Gunn}, J., et al.\ 2001, \aj, 122, 2850

\reference{}{Bell}, E.~F., {Wolf}, C., {Meisenheimer}, K., {Rix}, H., {Borch}, A., {Dye},  S., {Kleinheinrich}, M., et al.\ 2004, \apj, 608, 752

\reference{}{Bender}, R. 1988, \aap, 202, L5

\reference{}{Bernardi}, M., {Renzini}, A., {da Costa}, L.~N., {Wegner}, G., {Alonso},  M.~V., {Pellegrini}, P.~S., {Rit{\' e}}, C., \& {Willmer}, C.~N.~A. 1998,  \apjl, 508, L143

\reference{}{Bernardi}, M., {Sheth}, R.~K., {Annis}, J., {Burles}, S., {Finkbeiner}, D.~P.,  {Lupton}, R.~H., {Schlegel}, D.~J., et al.\ 2003, \aj, 125, 1882

\reference{}{Bertin}, E. \& {Arnouts}, S. 1996, \aaps, 117, 393

\reference{}{Binney}, J. 2004, \mnras, 347, 1093

\reference{} Binney, J., \& Tremaine, S. 1987, Galactic Dynamics
(Princeton: Princeton Univ.~Press)

\reference{}{Blanton}, M.~R., {Dalcanton}, J., {Eisenstein}, D., {Loveday}, J., {Strauss},  M.~A., {SubbaRao}, M., {Weinberg}, D.~H., et al.\ 2001, \aj, 121, 2358

\reference{}{Bower}, R.~G., {Kodama}, T., \& {Terlevich}, A. 1998, \mnras, 299, 1193

\reference{}{Bower}, R.~G., {Lucey}, J.~R., \& {Ellis}, R.~S. 1992, \mnras, 254, 601

\reference{}{Boylan-Kolchin}, M., {Ma}, C.-P., \& {Quataert}, E. 2005, {}MNRAS, submitted  (astro-ph/0502495)

\reference{}{Carlberg}, R.~G. 1986, \apj, 310, 593

\reference{}Ciotti, L., \& van Albada, T.~S. 2001, \apj, 552, L13

\reference{}{Colbert}, J.~W., Mulchaey, J.~S., \& Zabludoff, A.~I. 2001,
AJ, 121, 808

\reference{}{Cole}, S., {Lacey}, C.~G., {Baugh}, C.~M., \& {Frenk}, C.~S. 2000, \mnras,  319, 168

\reference{}{Coleman}, G.~D., {Wu}, C.-C., \& {Weedman}, D.~W. 1980, \apjs, 43, 393

\reference{}{Combes}, F., {Rampazzo}, R., {Bonfanti}, P.~P., {Pringniel}, P., \&  {Sulentic}, J.~W. 1995, \aap, 297, 37

\reference{}{Conselice}, C.~J., {Bershady}, M.~A., {Dickinson}, M., \& {Papovich}, C. 2003,  \aj, 126, 1183

\reference{} Cooray, A., \& Milosavljevic, M. 2005, ApJL, submitted (astro-ph/0503596)

\reference{}{Daddi}, E., {Cimatti}, A., {Renzini}, A., {Vernet}, J., {Conselice}, C.,  {Pozzetti}, L., {Mignoli}, M., {Tozzi}, P., {et al.} 2004, \apjl, 600, L127

\reference{}{Davoust}, E. \& {Prugniel}, P. 1988, \aap, 201, L30

\reference{} Dekel, A., \& Birnboim, Y. 2005, preprint (astro-ph/0412300)

\reference{} Dey, A., et al.\ 2005, submitted

\reference{}{Djorgovski}, S. \& {Davis}, M. 1987, \apj, 313, 59

\reference{}{Dubinski}, J., {Mihos}, J.~C., \& {Hernquist}, L. 1996, \apj, 462, 576

\reference{}{Eggen}, O.~J., {Lynden-Bell}, D., \& {Sandage}, A.~R. 1962, \apj, 136, 748

\reference{}{Ellis}, R.~S., {Smail}, I., {Dressler}, A., {Couch}, W.~J., {Oemler}, A.~J.,  {Butcher}, H., \& {Sharples}, R.~M. 1997, \apj, 483, 582

\reference{}{Ferrarese}, L. \& {Merritt}, D. 2000, \apjl, 539, L9

\reference{}{Franx}, M., {Illingworth}, G., \& {Heckman}, T. 1989, \aj, 98, 538

\reference{}{Franx}, M. \& {Illingworth}, G.~D. 1988, \apjl, 327, L55

\reference{}{Franx}, M., {Labb{\' e}}, I., {Rudnick}, G., {van Dokkum}, P.~G., {Daddi}, E.,  {F{\" o}rster Schreiber}, N.~M., {Moorwood}, A., {Rix}, H., {et al.} 2003, \apjl, 587, L79


\reference{} Fukugita, M., Shimasaku, K., Ichikawa, T. 1995, PASP, 107, 945

\reference{}{Fukugita}, M., {Hogan}, C.~J., \& {Peebles}, P.~J.~E. 1998, \apj, 503, 518

\reference{}{Gal}, R.~R., {de Carvalho}, R.~R., {Lopes}, P.~A.~A., {Djorgovski}, S.~G.,  {Brunner}, R.~J., {Mahabal}, A., \& {Odewahn}, S.~C. 2003, \aj, 125, 2064

\reference{}Gao, L., Loeb, A., Peebles, P.~J.~E., White, S.~D.~M., Jenkins,
A. 2004, ApJ, 614, 17

\reference{} Gawiser, E., et al.\ 2005, ApJS, submitted

\reference{}{Gebhardt}, K., {Bender}, R., {Bower}, G., {Dressler}, A., {Faber}, S.~M.,  {Filippenko}, A.~V., {Green}, R., {Grillmair}, C., {et al.} 2000, \apjl, 539, L13

\reference{}{Glazebrook}, K., {Abraham}, R.~G., {McCarthy}, P.~J., {Savaglio}, S., {Chen},  H., {Crampton}, D., {Murowinski}, R., {J{\o}rgensen}, I., {et al.} 2004, \nat, 430, 181

\reference{}{Gonz{\' a}lez-Garc{\'{\i}}a}, A.~C. \& {van Albada}, T.~S. 2003, \mnras, 342,  L36

\reference{}{Hernquist}, L. 1992, \apj, 400, 460

\reference{} Hernquist, L., \& Spergel, D.~N. 1995, ApJ, 399, L117

\reference{}{Hibbard}, J.~E. \& {van Gorkom}, J.~H. 1996, \aj, 111, 655

\reference{}{Holden}, B.~P., {van der Wel}, A., {Franx}, M., {Illingworth}, G.~D.,  {Blakeslee}, J.~P., {van Dokkum}, P., {Ford}, et al.\ 2005, \apjl, 620, L83

\reference{}{Hunsberger}, S.~D., {Charlton}, J.~C., \& {Zaritsky}, D. 1996, \apj, 462, 50

\reference{} Ibata, R., Irwin, M., Lewis, G., Ferguson, A.~M.~N.,
\& Tanvir, N. 2001, Nature, 412, 49

\reference{}{Jansen}, R.~A., {Franx}, M., {Fabricant}, D., \& {Caldwell}, N. 2000, \apjs,  126, 271


\reference{} Januzzi, B.~T., \& Dey, A. 1999, in ``Photometric
Redshifts and the Detection of High Redshift Galaxies'', ASP Conference
Series, Vol.\ 191, R.\ Weyman, L.\ Storrie-Lombardi, M.\ Sawicki, and
R. Brummer, Eds., p.\ 111

\reference{} Januzzi, B.~T., et al. 2005, submitted

\reference{}{Jimenez}, R., {Friaca}, A.~C.~S., {Dunlop}, J.~S., {Terlevich}, R.~J.,  {Peacock}, J.~A., \& {Nolan}, L.~A. 1999, \mnras, 305, L16

\reference{}{Johnston}, K.~V., {Sackett}, P.~D., \& {Bullock}, J.~S. 2001, \apj, 557, 137

\reference{}{J\o{}rgensen}, I., {Franx}, M., \& {Kj\ae{}rgaard}, P. 1995, \mnras, 273, 1097

\reference{}{J\o{}rgensen}, I., {Franx}, M., \& {Kj\ae{}rgaard}, P. 1996, \mnras, 280, 167

\reference{}{Kauffmann}, G. 1996, \mnras, 281, 487

\reference{}{Kauffmann}, G. \& {Haehnelt}, M. 2000, \mnras, 311, 576

\reference{}{Kauffmann}, G., {White}, S.~D.~M., \& {Guiderdoni}, B. 1993, \mnras, 264, 201

\reference{} Keres, D., Katz, N., Weinberg, D.~H., Dave, R. 2005,
MNRAS, submitted (astro-ph/0407095)

\reference{}{Khochfar}, S. \& {Burkert}, A. 2003, \apjl, 597, L117

\reference{} Lin, L., et al.\ 2004, ApJ, 617, L9

\reference{}{L{\' o}pez-Cruz}, O., {Barkhouse}, W.~A., \& {Yee}, H.~K.~C. 2004, \apj, 614,  679

\reference{}{Le F{\` e}vre}, O., {Abraham}, R., {Lilly}, S.~J., {Ellis}, R.~S.,  {Brinchmann}, J., {Schade}, D., {Tresse}, L., et al. 2000, \mnras, 311, 565

\reference{}{Makino}, J. \& {Hut}, P. 1997, \apj, 481, 83

\reference{}{Malin}, D.~F. \& {Carter}, D. 1983, \apj, 274, 534

\reference{} McIntosh, D.~H., Bell, E.~F., Rix, H.-W., Wolf, C., Heymans, C.,
Peng, C.~Y., Somerville, R.~S., Barden, M. 2005, \apj, submitted
(astro-ph/0411772)


\reference{}{Meza}, A., {Navarro}, J.~F., {Steinmetz}, M., \& {Eke}, V.~R. 2003, \apj, 590,  619

\reference{}{Mihos}, J.~C. 1995, \apjl, 438, L75

\reference{}{Mirabel}, I.~F., {Dottori}, H., \& {Lutz}, D. 1992, \aap, 256, L19

\reference{}{Murali}, C., {Katz}, N., {Hernquist}, L., {Weinberg}, D.~H., \& {Dav{\' e}},  R. 2002, \apj, 571, 1

\reference{}{Nakamura}, O., {Fukugita}, M., {Yasuda}, N., {Loveday}, J., {Brinkmann}, J.,  {Schneider}, D.~P., {Shimasaku}, K., \& {SubbaRao}, M. 2003, \aj, 125, 1682

\reference{}{Ostriker}, J.~P. 1980, Comments on Astrophysics, 8, 177

\reference{} Patton, D.~R., Carlberg, R.~G., Marzke, R.~O., Pritchet, C.~J.,
da Costa, L.~N., \& Pellegrini, P.~S. 2000, ApJ, 536, 153

\reference{}{Patton}, D.~R., {Pritchet}, C.~J., {Carlberg}, R.~G., {Marzke}, R.~O., {Yee},  H.~K.~C., {Hall}, P.~B., {Lin}, H., et al.\ 2002, \apj, 565, 208

\reference{} Peebles, P.~J.~E. 2002,  in ``A new era in cosmology'',
ASP Conference
Series, Vol.\ 283, N.~Metcalfe and T.~Shanks, Eds., p.~351

\reference{}{Peletier}, R.~F., {Davies}, R.~L., {Illingworth}, G.~D., {Davis}, L.~E., \&  {Cawson}, M. 1990, \aj, 100, 1091

\reference{}{Pildis}, R.~A., {Bregman}, J.~N., \& {Schombert}, J.~M. 1995, \aj, 110, 1498

\reference{}{Rix}, H.~R. \& {White}, S.~D.~M. 1989, \mnras, 240, 941

\reference{} Sandage, A., \& Visvanathan, N. 1978, ApJ, 223, 707

\reference{}{Sanders}, D.~B., {Soifer}, B.~T., {Elias}, J.~H., {Madore}, B.~F., {Matthews},  K., {Neugebauer}, G., \& {Scoville}, N.~Z. 1988, \apj, 325, 74

\reference{}{Sansom}, A.~E., {Hibbard}, J.~E., \& {Schweizer}, F. 2000, \aj, 120, 1946

\reference{}{Schweizer}, F. 1982, \apj, 252, 455

\reference{}{Schweizer}, F., {Miller}, B.~W., {Whitmore}, B.~C., \& {Fall}, S.~M. 1996,  \aj, 112, 1839

\reference{} Schweizer, F., Seitzer, P., Faber, S.~M., Burstein, D., Dalle
 Ore, C.~M., Gonzalez, J.~J. 1990, ApJ, 364, L33

\reference{}{Schweizer}, F. \& {Seitzer}, P. 1992, \aj, 104, 1039

\reference{} Silva, D.~R., \& Bothun, G.~D. 1998, \aj, 116, 85

\reference{} Somerville, R.~S. 2004, in ``Coevolution of Black Holes and
Galaxies'', L.\ C.\ Ho, Ed., Cambridge University Press, p.\ 391

\reference{}{Somerville}, R.~S., {Primack}, J.~R., \& {Faber}, S.~M. 2001, \mnras, 320, 504

\reference{} Spergel, D.~N., et al.\ 2003, ApJS, 148, 175

\reference{} Strateva, I., et al.\ 2001, AJ, 122, 1861

\reference{}{Tonry}, J.~L., {Howell}, S.~B., {Everett}, M.~E., {Rodney}, S.~A., {Willman},  M., \& {VanOutryve}, C. 2005, \pasp, 117, 281

\reference{}{Toomre}, A. \& {Toomre}, J. 1972, \apj, 178, 623

\reference{} Trager, S.~C., Faber, S.~M., Worthey, G., \& Gonz\'alez, J.~J.
2000, \aj, 119, 1645

\reference{} Tran, K.-V., van Dokkum, P.~G., Franx, M., Illingworth,
G.~D., Kelson, D.~D., F\"orster Schreiber, N. 2005, ApJ, 627, L25

\reference{}{Treu}, T., {Stiavelli}, M., {Casertano}, S., {Moller}, P., \& {Bertin}, G.  1999, \mnras, 308, 1037

\reference{} ---. 2002, \apjl, 564, L13

\reference{} Treu, T., Ellis, R.~S., Liao, T.~X., \& van Dokkum, P.~G.
2005, \apj, 622, L5

\reference{} van der Marel, R.~P. 1991, MNRAS, 253, 710

\reference{} van der Wel, A., Franx, M., van Dokkum, P.~G., Rix, H.-W.,
Illingworth, G.~D., \& Rosati, P. 2005, \apj, in press (astro-ph/0502228)

\reference{}{van Dokkum}, P.~G. \& {Franx}, M. 1995, \aj, 110, 2027

\reference{} ---. 2001, \apj, 553, 90

\reference{}{van Dokkum}, P.~G., {Franx}, M., {Fabricant}, D., {Kelson}, D.~D., \&  {Illingworth}, G.~D. 1999, \apjl, 520, L95

\reference{}{van Dokkum}, P.~G., {Franx}, M., {Kelson}, D.~D., \& {Illingworth}, G.~D.  1998a, \apjl, 504, L17

\reference{} ---. 2001a, \apjl, 553, L39

\reference{}{van Dokkum}, P.~G., {Franx}, M., {Kelson}, D.~D., {Illingworth}, G.~D.,  {Fisher}, D., \& {Fabricant}, D. 1998b, \apj, 500, 714

\reference{}{van Dokkum}, P.~G., {Stanford}, S.~A., {Holden}, B.~P., {Eisenhardt}, P.~R.,  {Dickinson}, M., \& {Elston}, R. 2001b, \apjl, 552, L101

\reference{}{White}, S.~D.~M. \& {Frenk}, C.~S. 1991, \apj, 379, 52

\reference{} Wyithe, J.~S.~B., \& Loeb, A. 2005, ApJ, submitted
(astro-ph/0506294)
\reference{} Wolf, C., et al.\ 2005, \aap, submitted

\reference{}{Yi}, S.~K., {Yoon}, S.-J., {Kaviraj}, S., {Deharveng}, J.-M., {Rich}, R.~M.,  {Salim}, S., {Boselli}, A., et al.\ 2005, \apjl, 619, L111

\reference{}{York}, D.~G., {Adelman}, J., {Anderson}, J.~E., {Anderson}, S.~F., {Annis},  J., {Bahcall}, N.~A., {Bakken}, J.~A., et al.\ 2000, \aj, 120, 1579

\reference{}{Zabludoff}, A.~I. \& {Mulchaey}, J.~S. 1998, \apj, 496, 39

\reference{}{Zepf}, S.~E., {Whitmore}, B.~C., \& {Levison}, H.~F. 1991, \apj, 383, 524

\end{references}


\begin{appendix}
\section{Catalog and Atlas}
Here we provide coordinates, magnitudes, colors, morphologies,
and tidal classifications for all 126 red galaxies.
We also present images of all objects in the sample. 
These small images at fixed contrast level
do not do the data justice, but space
limitations prohibit an atlas on the scale of Figs.\
\ref{example_col.plot} and \ref{example_bw.plot}.
\vspace{1cm}\\
{\large  The atlas is available at
www.astro.yale.edu/dokkum/mergers/ along with a version of the
paper with {\bf much} higher quality figures.
}
\end{appendix}

\end{document}